\documentclass[10pt,aps,prl,twocolumn,superscriptaddress]{revtex4-2}
\usepackage[utf8]{inputenc}
\usepackage{amsmath}
\usepackage{amssymb}
\usepackage{graphicx}
\usepackage{hyperref}
\usepackage{xcolor}
\usepackage[caption=false]{subfig}

\newcommand{\ket}[1]{\lvert #1 \rangle}
\newcommand{\bra}[1]{\langle #1 \rvert}

\newcommand{\norm}[1]{\left\lVert #1 \right\rVert}

\def\be{\begin{equation}}      
\def\ee{\end{equation}}
\def\beu{\begin{equation*}}   
\def\eeu{\end{equation*}}

\providecommand{\mean}[1]{\langle#1\rangle}

\definecolor{new}{rgb}{.08,.05,.8}

\newcommand{\delete}[1]{{}} 

\usepackage[normalem]{ulem}

\begin{document}
\title{Operator scaling dimensions and multifractality at measurement-induced transitions}
\date{\today}

\author{A. Zabalo}
\affiliation{Department of Physics and Astronomy, Center for Materials Theory, Rutgers University, Piscataway, New Jersey 08854, USA}

\author{M. J. Gullans}
\affiliation{Department of Physics, Princeton University, Princeton, New Jersey 08544, USA}
\affiliation{Joint Center for Quantum Information and Computer Science, NIST/University of Maryland, College Park, Maryland 20742, USA}

\author{J. H. Wilson}
\affiliation{Department of Physics and Astronomy, Center for Materials Theory, Rutgers University, Piscataway, New Jersey 08854, USA}
\affiliation{Department of Physics and Astronomy, Louisiana State University, Baton Rouge, LA 70803, USA}
\affiliation{Center for Computation and Technology, Louisiana State University, Baton Rouge, LA 70803, USA}

\author{R. Vasseur}
\affiliation{Department of Physics, University of Massachusetts, Amherst, Massachusetts 01003, USA}

\author{A.~W.~W.~Ludwig}
\affiliation{Department of Physics, University of California, Santa Barbara, California 93106, USA}

\author{S. Gopalakrishnan}
\affiliation{Department of Physics, The Pennsylvania State University, University Park, Pennsylvania 16802, USA}
\affiliation{Department of Physics and Astronomy, CUNY College of Staten Island, Staten Island, New York 10314, USA}

\author{David A. Huse}
\affiliation{Department of Physics, Princeton University, Princeton, New Jersey 08544, USA}

\author{J. H. Pixley}
\affiliation{Department of Physics and Astronomy, Center for Materials Theory, Rutgers University, Piscataway, New Jersey 08854, USA}
 \affiliation{Department of Physics, Princeton University, Princeton, New Jersey 08544, USA}
 \affiliation{Center for Computational Quantum Physics, Flatiron Institute, 162 5th Avenue, New York, New York 10010}

\begin{abstract}
Repeated local measurements of quantum many-body systems can induce a phase transition in their entanglement structure.
These measurement-induced phase transitions (MIPTs) have been studied for various types of dynamics, yet most cases yield quantitatively similar 
critical exponents, making it unclear how many distinct
universality classes are present.
Here, we 
probe the properties of the conformal field theories governing these MIPTs  using a numerical transfer-matrix method, which allows us to extract the effective central charge, as well as the first few low-lying scaling dimensions of operators at these critical points for $(1+1)$-dimensional systems.
Our results provide convincing evidence that the generic and Clifford MIPTs for qubits lie in different universality classes and that both are distinct from the percolation transition for qudits in the limit of large on-site Hilbert space dimension.
For the generic case, we find strong evidence of multifractal scaling of correlation functions at the critical point, reflected in a continuous spectrum of scaling dimensions.
\end{abstract}

\maketitle

The dynamics of an open quantum system can be viewed as unitary evolution
interspersed with events where an environment \emph{measures} the system. This
competition between entangling dynamics and collapsing measurements
leads to a measurement-induced phase transition (MIPT) between phases with
distinct entanglement structure~\cite{skinner2019measurement,PhysRevB.98.205136,PhysRevB.99.224307,li2019measurement,Gullans20,PhysRevLett.125.030505,
zabalo2020critical,jian2020measurement,bao2020theory,gullans2020scalable,li2020conformal,szyniszewski2020universality,Fan20}.
By increasing the frequency of measurements, the system goes from a volume-law phase where the entanglement entropy of a subsystem scales with its volume
to an area-law phase where it scales with its boundary. This transition occurs
in the individual ``trajectories'' but is invisible in the mixed state averaged over measurement outcomes.

MIPTs exist in various classes of dynamics~\cite{Lavasani21,Sang20,ippoliti2021entanglement,fuji2020measurement,alberton2021entanglement,lang2020entanglement,chen2020emergent,lunt2020measurement,nahum2020entanglement,Jian2020CriticalityNonUnitary,jian2021syk,bentsen2021measurement,nahum2021measurement,gopalakrishnan2021entanglement}, have been observed experimentally~\cite{Noel21exp}, and are analytically tractable in certain limits, interpreted as a
percolation transition~\cite{skinner2019measurement,jian2020measurement,bao2020theory}.
Even away from tractable limits, the numerically extracted critical exponents of the MIPT are close to the values for percolation~\cite{zabalo2020critical}.
These observations raise the question: Are MIPTs resulting from different dynamics in distinct universality classes?

Beyond classifying the universal nature of MIPTs, developing precise characterizations of this class of critical phenomena has motivations in quantum information and computational complexity theory.
In particular, an entanglement transition potentially signifies a phase transition in the resources required to represent the quantum state on a classical computer~\cite{napp2019efficient,noh2020efficient}.
Such quantum information-theoretic observables lack natural counterparts in the conventional framework of statistical physics.
Consequently, our understanding of the ``relevant'' degrees of freedom in describing the related critical phenomena remains nascent.

This work presents evidence that MIPTs in different classes of random circuits belong to distinct universality classes beyond percolation.
These conclusions are supported by a numerical exploration of the non-unitary conformal field theories (CFTs) with central charge $c=0$ governing the MIPTs for three classes of dynamics---
generic (Haar), dual-unitary, and Clifford random circuits, each with random single-site measurements of Pauli operators.
The emergence of conformal invariance at MIPTs is suggested by mappings onto statistical models~\cite{RTN2,jian2020measurement,bao2020theory} and confirmed in previous 
numerical work~\cite{li2020conformal}.
We probe the properties of these CFTs by numerically computing several leading Lyapunov exponents of the transfer matrix. The Lyapunov exponents are related to the
scaling dimensions characterizing the scaling of {\it typical}~\cite{LudwigHierarchiesNPB1990} 
observables of the CFT, the first of which is related to the
``effective central charge'' 
$c_\mathrm{eff}$
~\cite{LudwigCardyNPB1987} -- a universal 
number~\footnote{{\it different from} the prefactor of the log  of subsystem size in the entanglement entropy which is instead a (boundary) scaling dimension~\cite{RTN2,jian2020measurement}}
distinguishing CFTs with central charge $c=0$.

We find evidence that the MIPT for generic circuits
belongs to a different universality class than that for Clifford circuits, while both differ from percolation. 
The effective central charge is distinct in the two cases: $c_\mathrm{eff}^H \approx 0.25(3)$ and $c_\mathrm{eff}^C\approx 0.37(1)$, respectively. We  compare these numerical values to the predictions of  large on-site Hilbert space ($d \to \infty$) mappings onto percolation:
$c_\mathrm{eff}^{H,d \to \infty} \approx 0.291$ for Haar and  $c_\mathrm{eff}^{C,d=2^n \to \infty} \approx 0.365$ for Clifford qudit circuits.
Dual-unitary circuits have a transition in the generic universality class, but their symmetries allow us to extract the 
 effective central charge 
 $c^\mathrm{DU}_\mathrm{eff} = 0.24(2)$ and the leading
Lyapunov exponents
with higher precision. 
We also find evidence that the spectra of operators at MIPTs are distinct from those in the percolation CFT. Thus the generic and Clifford MIPTs appear to be governed by two distinct CFTs and differ from any previously known instances. 
Last, we demonstrate  multifractality in the generic MIPT in a chain of qubits. 

\begin{figure*}[tb]
    \centering
    \includegraphics[width=\linewidth]{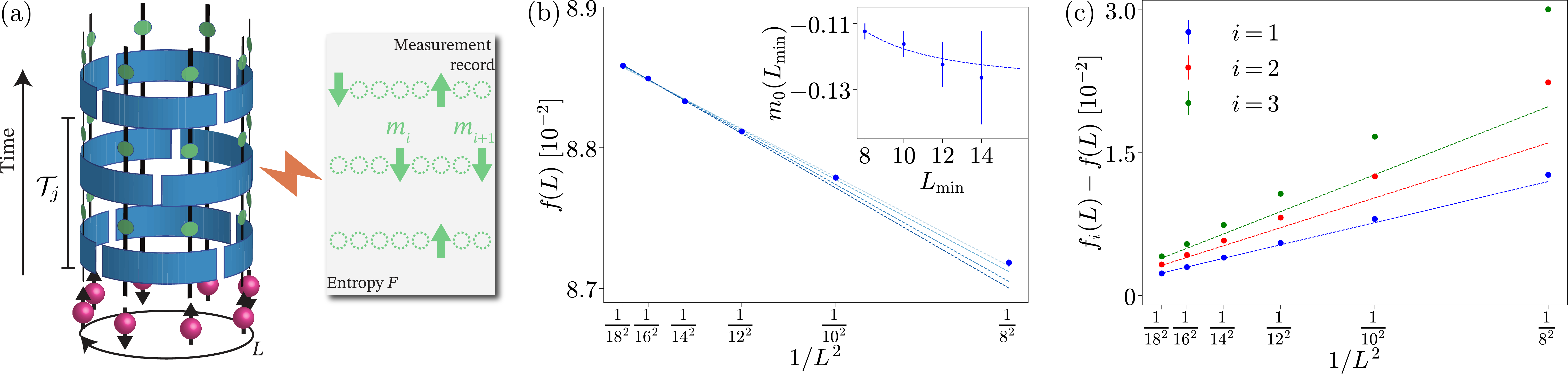}
    \caption{\label{fig:cyl_f0_fif0}
    (a) The cylindrical geometry of the random circuit model for a
        system of qubits of length $L$ with periodic boundary conditions. The blue rectangles
        represent the entangling gates and the green dots are the location of
        measurements. The time evolution can be viewed as a product of transfer
        matrices denoted by $\mathcal{T}_j$ whose leading Lyapunov exponent is given by the entropy of the measurement record $F$.
        (b) The free energy density displays the $1/L^2$
        dependence expected from Eq.~\eqref{eq:ceff}
        allowing us to extract $c_{\mathrm{eff}}$. Darker
        blue indicates increasing $L_\mathrm{min} = 8\to 14$.
        (Inset) 
        To improve our estimate  we
        successively remove the smallest system size from the fit and find
        $m_0(L_\mathrm{min})$ which contains the leading order correction to
        Eq.~\eqref{eq:ceff}. The dotted line corresponds to the fit
        $m_0(L)=-0.13 + \frac{0.98}{L^2}$.
        (c) The differences of the
        generalized free energy densities, $f_i(L) - f(L)$,
        shows the $1/L^2$ behavior expected from
        Eq.~\eqref{eqn:higherFs}. The dotted lines correspond to the extrapolated
        values $m_i(L_\mathrm{min}\to\infty)$.
        The data shown is for the dual unitary model at $p=p_c\approx 0.14$ and $25\,000$ samples for $L=8,10,12,14,16$ and $
        10\,000$ for $L = 18$.
        }
\end{figure*}

\emph{From quantum channels to CFTs}.---
Consider a quantum circuit with a fixed set of unitary gates and measurement locations and times. 
The hybrid unitary/measurement dynamics is described through the quantum channel
\begin{equation}\label{krausdec}
  \mathcal{N}_{t}(\rho) = \sum_{\mathbf{m}} K_{\mathbf{m}} \rho K^{\dagger}_{\mathbf{m}},
\end{equation}
where $\rho$ is the system's density matrix, and
$K_{\mathbf{m}} = K_t^{m_t} K_{t-1}^{m_{t-1}} \ldots K_1^{m_1}$ is a Kraus operator. 
The operators 
$K_s^{m_s} = P_s^{m_s} U_s$ consist of random unitary gates $U_s$ and random projectors $P_s^{m_s}$ onto measurement outcomes $m_s$.
Each summand $K_\mathbf{m} \rho K_\mathbf{m}^\dagger$ in Eq.~\eqref{krausdec} represents a ``quantum trajectory'' of the system.
Moreover, $\mathrm{Tr}(K_\mathbf{m} \rho K_\mathbf{m}^\dagger) = p_{\mathbf{m}} (\rho)$, is the probability of the set of outcomes $\mathbf{m}$. 
We suppress the argument $\rho$ since at late times the probabilities $p_\mathbf{m}$ become independent of the initial density matrix at the critical point. 

Following Ref.~\cite{li2020conformal}, we posit that each trajectory can be identified with a $(1+1)$-dimensional statistical mechanics model, 
defined implicitly through the identification that its partition function $Z_\mathbf{m} \equiv p_\mathbf{m}$.
Without defining an explicit model, we note that the partition functions of canonical statistical mechanics models can be written as tensor networks with a similar structure to the single-trajectory circuit~\cite{PhysRevLett.99.120601}, so this identification is natural. 
The trajectories making up a particular channel form an ensemble of statistical mechanics models with quenched spacetime randomness due to the measurement outcomes. 
Each model's weight in the ensemble is set by its Born probability $p_\mathbf{m}$. 

It follows from these observations that, for a circuit of fixed length $L$, a layer of time evolution for a particular trajectory (i.e., the map $\rho \to \mathcal{T}_t \rho \mathcal{T}_t^\dagger$, where $\mathcal{T}_j=K^{m_{2j}}_{2j}K^{m_{2j-1}}_{2j-1}$ is depicted in Fig.~\ref{fig:cyl_f0_fif0}a) acts as a transfer matrix for the statistical mechanics model describing that trajectory. Note that one can write $Z_\mathbf{m} = \sum_i  (\sigma^{\mathbf{m}}_i)^2$, where $(\sigma^{\mathbf{m}}_i)^2$ are the eigenvalues of $K_{\mathbf{m}} K_{\mathbf{m}}^\dagger$, i.e., the squares of the singular values of $K_\mathbf{m}$.  Equivalently, these are the eigenvalues of an initially completely mixed density matrix that is purified by the evolution \cite{Gullans20}.
At late times, 
$K_{\bf m}$ is given by a large product of the operators $\mathcal{T}_j$ and 
$\sigma^{\mathbf{m}}_i$ decays exponentially, as the state purifies. This exponential decay motivates the definition of trajectory dependent
exponents~\cite{LudwigHierarchiesNPB1990,jacobsen1998critical} $\lambda_i^{\mathbf m}$, through $\left[\sigma_i^{\mathbf m}(t)\right]^2 = e^{\lambda_i^{\mathbf m}t}$ as $t\rightarrow\infty$; note that $\lambda_i^{\mathbf m}<0$, and we compute them as specified in~\cite{supp}. 
We then average $\lambda_i^{\mathbf m}$ over trajectories (using the Born weights $p_{\mathbf m}$) to yield the Lyapunov exponents $\lambda_0,\lambda_1,\lambda_2,\dots$ in descending order.

The leading Lyapunov exponent of the transfer matrix has an appealing interpretation.
In general, this quantity is 
the free energy of the statistical mechanics model up to a factor of time, i.e., $t \lambda_0^{{\bf m}}=\ln p_{\bf m}$. 
Averaging the free energy with Born weights gives us that $F/t=-\lambda_0$ where
\begin{equation}
F = -\sum_\mathbf{m} p_\mathbf{m} \ln p_\mathbf{m}.
\label{eqn:F}
\end{equation}
This averaged free energy is the Shannon entropy of the measurement record, see Fig.~\ref{fig:cyl_f0_fif0}a.

As in more conventional disordered systems, the averaged free energy can be computed within a replica formalism. Introducing the annealed average replicated partition function $\bar{Z}_r = \sum_{\mathbf{m}} p_{\mathbf{m}} Z_{\mathbf{m}}^r$ where $r$ is the replica index, the corresponding annealed average free energy is $F_r = -\ln \bar{Z}_r$. The quenched average free energy from Eq.~\eqref{eqn:F} is then given by 
$F =  \lim_{r \to 0}\frac{d F_r}{dr}$ in the replica limit $r \to 0$. The  annealed average replicated statistical model has a phase transition for finite $r>0$, which we assume is described by a CFT whose properties approach those of the MIPT in the $r \to 0$ limit.

\emph{Effective central charge and operator spectrum}.---
The central charge $c(r)$ of the CFT describing the replicated model $\bar{Z}_r$ goes to $c(r) \to 0$ in the replica limit $r\rightarrow 0$; this follows from the trivial partition function $\bar{Z}_{r \to 0} =1$. 
However, standard CFT results on a cylinder of circumference $L$ and length $t$ (in the limit $t\gg L$)
imply that the averaged free energy density $F(L,t)/A=f(L)$~\cite{LudwigCardyNPB1987,jacobsen1998critical} scales as
\begin{equation}
f(L) = f(L = \infty) - \frac{\pi c_{\mathrm{eff}}}{6L^2} + \ldots
\label{eq:ceff}
\end{equation}
where $c_{\mathrm{eff}} = \lim_{r \to 0}\frac{d c(r)}{dr}$ is a universal number called the effective central charge, and $A \equiv \alpha L t$ is the effective spacetime area. Since the statistical mechanics model is only defined implicitly, its intrinsic space and time scales (and the anisotropy $\alpha$ between them) must be extracted numerically, as we discuss below. 

We now turn to the subleading Lyapunov exponents. In the statistical mechanics picture, the difference of the two leading Lyapunov exponents controls the decay of correlations along the direction of the transfer matrix, i.e., it determines the scale on which initial conditions are forgotten. The next-to-leading Lyapunov exponent thus corresponds to the most relevant (i.e., longest-lived) operator while higher Lyapunov exponents correspond to faster-decaying operators. Conformal invariance dictates how these quantities behave at critical points:
\begin{equation}
f_i(L) - f(L) = 2\pi x^{\rm typ}_i /L^2,
\label{eqn:higherFs}
\end{equation}
where $f_i(L)=-\lambda_i/(\alpha L)$ is obtained from the Lyapunov exponents $(i=1,2,\dots$) and $x^{\rm typ}_i$ is the scaling dimension of the $i^{\mathrm{th}}$ most relevant operator characterizing the decay of {\em typical}~\cite{LudwigHierarchiesNPB1990} correlators, defined only in the generic case --- averaged correlators will be discussed below. 

\emph{Circuit models}.---
We consider two main ensembles of random circuits: Haar random circuits
with two-qubit gates chosen from the Haar measure and stabilizer circuits
with gates chosen from the Clifford group. 
Stabilizer circuits have an efficient classical algorithm for the simulation of the single-circuit observables studied in this work \cite{Aaronson04}.
Additionally, we consider subclasses of Haar and Clifford circuits in which all gates are ``dual-unitary''~\cite{bertini2019exact, gopalakrishnan2019unitary}, i.e., unitary in both space
and time directions. The most generic dual unitary gates are given
by $U = e^{i\phi}(u_+ \otimes u_-) \cdot V[J] \cdot (v_- \otimes v_+)$, where $\phi, J \in \mathbb{R}, u_\pm, v_\pm \in \mathrm{SU}(2)$, and $V[J] = \exp[-i\left(\frac{\pi}{4}\sigma^x \otimes \sigma^x + \frac{\pi}{4}\sigma^y \otimes \sigma^y + J\sigma^z \otimes \sigma^z\right)]$~\cite{bertini2019exact}.
We present evidence that the dual unitary Haar (Clifford) circuits lie within the same universality class as Haar (Clifford) circuits (to within our numerical precision, see below). However, these circuits allow for a more accurate
estimate of the critical properties since their statistical self-duality under spacetime rotations forces $\alpha = 1$ and the associated rescaling factors are known exactly~\cite{supp}.
Below, all results are taken at the critical point determined
using the ancilla order parameter described in Ref.~\cite{gullans2020scalable}.
We find $p_c^H = 0.17(1), p_c^{DU} = 0.14(1), p_c^C = 0.1596(3)$ and $p_c^{DC}
= 0.205(1)$ for the Haar, dual unitary, Clifford, and dual Clifford models,
respectively~\cite{zabalo2020critical,supp}.

The anisotropy parameters for the Haar and Clifford models are estimated by
comparing the correlation functions 
along the space and time directions. These
correlation functions are determined in the quantum circuit by computing the mutual information
between two ancilla qubits separated in space and
time~\cite{gullans2020scalable,zabalo2020critical}. In the Haar model, 
$\alpha^\mathrm{H} = 0.81(9)$
while for the stabilizer model
$\alpha^\mathrm{C} = 0.62(3)$.
As a check, we compute the anisotropy for the dual unitary variants
and find $\alpha^\mathrm{DU}=1.0(1)$
in agreement with the known value 
$\alpha = 1$~\cite{supp}.

\emph{Numerical Approach}.---We now discuss our algorithm for finding the leading Lyapunov exponents in the Haar and dual unitary models (see~\cite{supp} for the approach used for Clifford and percolation models). 
The first few singular values $\sigma_i^{\bf m}(t)$ are computed by
picking a random initial state, generating a set of mutually orthogonal vectors to the initial state, and
iteratively applying the same set of transfer matrices $\mathcal{T}_j$ (depicted in Fig.~\ref{fig:cyl_f0_fif0}a) to the set. 
Each projector in $\mathcal{T}_j$ is chosen based on the Born probability of the time-evolved initial state and
after each application of $\mathcal{T}_j$ the set is re-orthogonalized. 
This allow us to estimate $F$ in Eq.~\eqref{eqn:F} and $f_i(L)=-\lambda_i/(\alpha L)$ in Eq.~\eqref{eqn:higherFs}~\cite{supp} through a Monte Carlo sampling of the Born probabilities~\cite{supp}.
We note that our results are sensitive to the initial state at early times; to achieve results independent of initial conditions, we wait an ``equilibration'' time of $\tau = 4L$ and average over different initial states (see supplement~\cite{supp}).
This approach agrees well with a direct evaluation of the spectrum of the transfer matrix on small system sizes~\cite{supp}.

\emph{Results.} ---
The data for the leading Lyapunov exponent at long times provides an estimate of 
$F(t\to\infty)$ 
and is shown in Fig.~\ref{fig:cyl_f0_fif0}b. We find that this displays
a clear linear behavior as a function of $1/L^2$ with slope $m_0$ related
to the effective central charge as expected from
Eq.~\eqref{eq:ceff}. To improve our estimate of $m_0$, we can
successively remove smaller system sizes, $L < L_\mathrm{min}$, from the fit
and write $m_0(L_\mathrm{min}) = m_0(\infty) + \frac{b}{L_\mathrm{min}^2}$
which accounts for the leading order correction to Eq.~\eqref{eq:ceff}. The
procedure is illustrated in Fig.~\ref{fig:cyl_f0_fif0}b and its inset. Using
$c_\mathrm{eff} = -\frac{6m_0(\infty)}{\pi}$, we find $c^\mathrm{H}_\mathrm{eff} =
0.25(3)$ for the Haar model
with an improved estimate of $c^\mathrm{DU}_\mathrm{eff} = 0.24(2)$
from the dual unitary variant.
Similarly, $c^\mathrm{C}_\mathrm{eff} = 0.37(1)$ for the stabilizer circuit~\cite{supp}.
A rudimentary analysis of $c_{\mathrm{eff}}$ as a function of $p$ displays a
broad maximum near $p_c$ suggesting deviations within the uncertainty of $p_c$
should not significantly affect the quoted values (results not shown).
These values can be compared to the exact predictions for large onsite Hilbert space dimension $d\to \infty$, where the MIPT maps onto percolation. Following methods developed in prior work \cite{RTN1,RTN2,PhysRevB.99.174205,jian2020measurement,bao2020theory}, we find $c_\mathrm{eff}^{H,d \to \infty} =\frac{5\sqrt{3}(1-\gamma)}{4\pi} = 0.291\dots$ in the Haar case and, using additional properties of the Clifford group proved in Ref.~\cite{2017arXiv171208628G}, $c_\mathrm{eff}^{C,d=2^n \to \infty} = 0.365\dots$ for stabilizer circuits~\cite{supp}. Our numerical estimate of $c^\mathrm{C}_\mathrm{eff}$ for qubits ($d=2$) is  consistent with the percolation value ($d=2^n\to \infty$), 
thus more exponents (or universal data) are needed to distinguish those two universality classes. 


\begin{table}[b!]
     \resizebox{\columnwidth}{!}{%
     \begin{tabular}{ccccc|c}
        \hline
        \hline
         & Haar & Dual  & Clifford & Dual & $d = \infty$ \\
         & & Unitary & &  Clifford & Haar/Clifford   \\
        \hline
         $c_{\rm eff}$ & 0.25(3) & 0.24(2) & 0.37(1) & &0.2914/0.3652   \\
         \hline
         $x_1$ & 0.14(2)$^\dagger$ & 0.122(1)$^\dagger$ & 0.120(5) & 0.111(1) & 0.1042 \\
                  \hline
         MF & \checkmark & \checkmark & $\times$ & $\times$ & $\times$ \\
        \hline
        \hline
    \end{tabular}}
\caption{\label{tbl:results}Critical data for the various models: effective central charge $c_{\rm eff}$, order-parameter exponent $x_1$, and whether order-parameter correlations exhibit multifractality (MF). For critical points exhibiting multifractality, we have quoted the order-parameter exponent governing \emph{typical} correlations (marked with $^\dagger$). This is not strictly comparable to the exponent governing \emph{average} correlations quoted for the three other models.}
\end{table}

The differences between Lyapunov exponents,
$f_i(L) - f(L) \sim 1/L^{2}$, as expected (Fig.~\ref{fig:cyl_f0_fif0}c); the slope of the fitted line, $m_i(L_\mathrm{min})$ can
then be used to determine $x^{\rm typ}_i$.
The scaling dimension $x^{\rm typ}_1$ is related to the (typical)
bulk exponent of the `order parameter', $x^{\rm typ}_1 = \eta/2$~\cite{gullans2020scalable}.
Our estimates for the Haar model $\eta^\mathrm{H}/2 = 0.14 \pm 0.02$
and the dual unitary variant $\eta^\mathrm{DU}/2 = 0.122 \pm 0.001$ are
consistent with the result $\eta/2 \approx 0.125$ for the mutual information computed in
Ref.~\cite{zabalo2020critical}, 
for Renyi indices $n>1$,
and are close to, but outside of error bars from, the percolation value $\eta/2 = \frac{5}{48} \approx 0.104$.
The next lowest scaling dimensions are given by $x^{\rm typ}_2 = 0.18(2)$ and $x^{\rm typ}_3 = 0.23(3)$
for the Haar model and $x^{\rm typ}_2 =  0.163(1)$ and $x^{\rm typ}_3 = 0.202(1)$ for the dual
unitary model.  It is unclear at present which operators these correspond to.
The error bars in $c_\mathrm{eff}$ and $x_i^\mathrm{typ}$ only include the uncertainty in the averaged measurement record (estimated via bootstrapping) and $\alpha$ as discussed in the supplement~\cite{supp}.

In the stabilizer circuit models, we have also extracted the order parameter exponent using an improved numerical method with the results given in Table~\ref{tbl:results}. Further details are provided in the supplement \cite{supp}, where we also generalize the order parameter exponent to an infinite hierarchy of ``purification'' exponents with distinct behavior from the minimal-cut percolation model.  We further improve our precision in extracting the order parameter exponent by using a dual-unitary Clifford model, where each two-qubit gate is drawn randomly from the uniform set of dual-unitary Clifford gates.  The critical $p_c=0.205(1)$ of this model violates a conjectured bound on $p_c \le 0.1893$ in 1+1-dimensions arising from the Hashing bound for the depolarizing channel  \cite{Fan20}.  In this dual-unitary Clifford model, we observe a significant  difference from the percolation value for the order parameter exponent, providing
convincing evidence that these models lie in different universality classes.  

\begin{figure}
    \centering
    \includegraphics[width=0.9\linewidth]{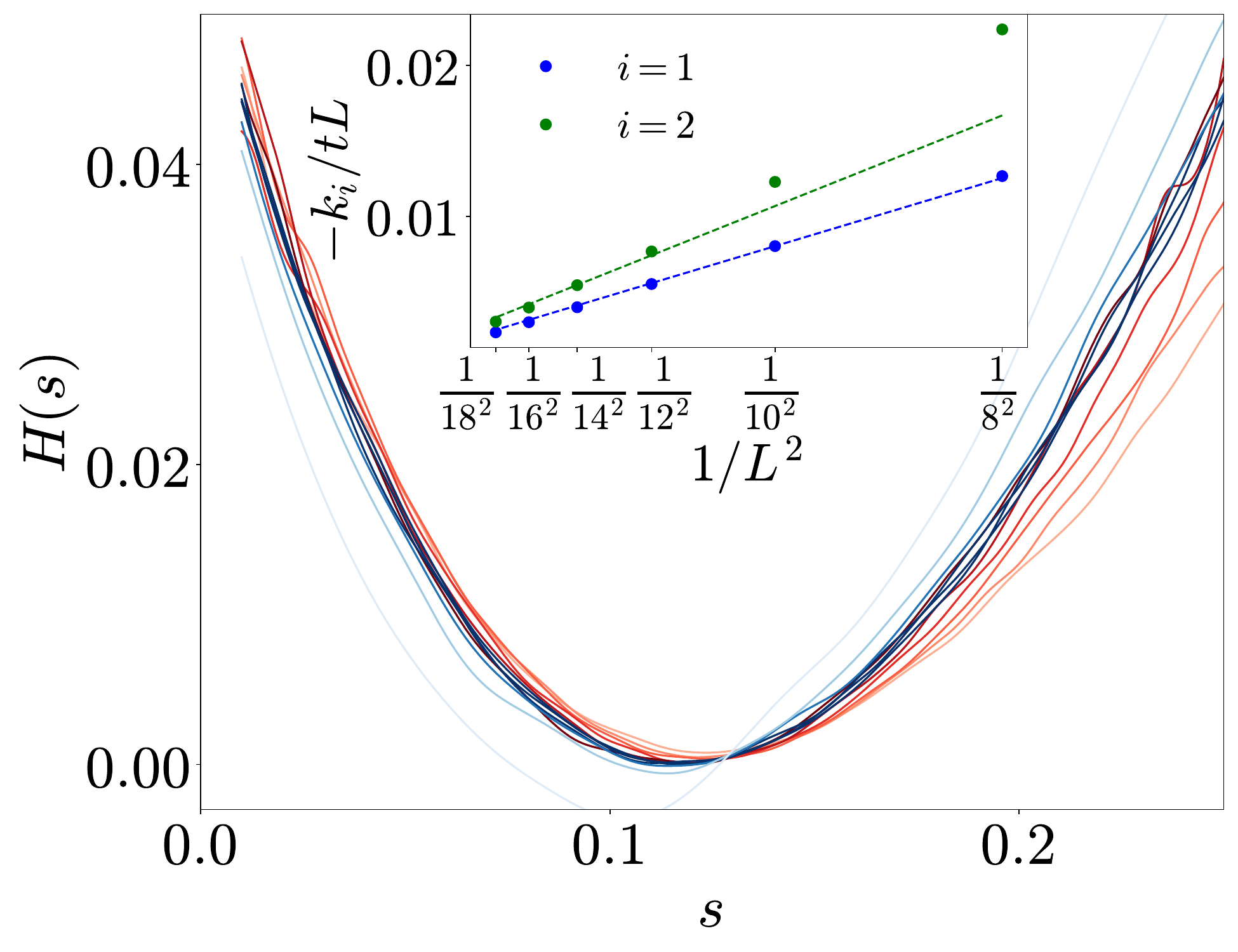} \\
    \caption{\label{fig:Halpha}The scaling collapse of the data onto a universal multifractal scaling
    function $H(s)$, given by Eq.~\eqref{eq:Halpha}, demonstrates multifractality at the
    critical point of the Haar transition and corresponds to a continuum of critical exponents.
    Data is shown for the dual unitary model (a similar quality of collapse also holds for Haar gates~\cite{supp})  where darker red indicates larger system sizes
    ($L = 8\to 18, t = 24 L$) and darker blue indicates later times ($t = 3L \to 24L, L = 16$).
    (inset) The first two cumulants $k_i$ of $\ln G_1(t)$
    divided by $tL$ show
    the expected $1/L^2$ behavior.
    }
    \end{figure}
    
\emph{Multifractality.}---The exponent $x_1^{{\rm typ}}$ captures how the correlation function 
of the order parameter, $G^{\mathbf{m}}_1(t)$---defined through 
$\ln G^{\mathbf{m}}_1(t)=t(\lambda_1^{{\bf m}}-\lambda_0^{{\bf m}})$
---
decays as $t \to \infty$ in a \emph{typical} trajectory $\mathbf{m}$. 
Specifically, 
$\overline{\ln G_1^{\bf m}(t)} \sim -(2\pi t/L) x_1^{\mathrm{typ}}$, when $t \gg L$, 
see Eq.~\eqref{eqn:higherFs}, where $\overline{\left(\ldots\right)}$ denotes an average over trajectories. Below, we suppress the trajectory index $\mathbf{m}$.
Quantities such as $\ln G_1(t)$ are self-averaging and can be extracted numerically. However, the decay of the \emph{sample-averaged} correlation function $\overline{G_1(t)}$ and its moments,
$\overline{G_1(t)^n}\sim \exp\left[-2\pi t x_{1}(n)/L\right]$  (in the limit 
$t \gg L$), are governed by a continuous family of critical exponents $x_1(n)$ due to multifractal scaling at the critical point of the Haar transition. 
We characterize the multifractal scaling through the distribution function $P[Y(t)]$ where $Y(t) \equiv - \ln G_1(t)$. If this correlation function exhibits multifractal scaling, its distribution will
follow the universal  scaling form~\cite{LudwigHierarchiesNPB1990}
\begin{equation}
    P\left[Y(t)\right] \sim \left( \frac{2\pi\alpha t}{L}\right)^{-1/2}\exp\left[-\frac{2\pi\alpha t}{L} H\left(\frac{Y(t)}{2\pi\alpha t/L}\right) \right],
    \label{eq:Halpha}
\end{equation}
for some (universal) function $H(s)$.
As shown in  Fig.~\ref{fig:Halpha}, our numerical results
for various system sizes and times, when rescaled according to Eq.~\eqref{eq:Halpha}
collapse onto a single curve, 
demonstrating multifractality at the Haar critical point. This observation is one of the central results of our work.

Finally, the exponents $x_1(n)$
are connected to the scaling function $H(s)$; one can use the standard relation between moments and cumulants
\begin{equation}
\ln\overline{G_1(t)^n} = n\overline{\ln G_1(t)} + \frac{n^2}{2!}\overline{\left(\ln G_1(t) - \overline{\ln G_1(t)}\right)^2} + \ldots,
\label{eq:lnG}
\end{equation}
where all terms are self-averaging, 
to find an expansion for the $n$th moment exponent $x_1(n) = n x_1^{(1)} + \frac{n^2}{2!} x_1^{(2)} + \ldots$, valid at small $n$. (Here, $x_1^{(1)} = x^{\rm typ}_1$.) In the inset of Fig.~\ref{fig:Halpha}, we see
that the first two cumulants 
$k_{1,2}$  of $\ln G_1(t)$  have, when divided by $tL$,
the expected $\sim 1/L^2$ scaling. We estimate $x_1^{(1)} = 0.14(2), x_1^{(2)} = 0.15(2)$ for the Haar model and $x_1^{(1)} = 0.122(1), x_1^{(2)} = 0.145(2)$ for the dual unitary model.
The substantial value of $x_1^{(2)}$ indicates that multifractality is strong: the average and typical exponents are appreciably different.

Concluding, we studied the effective central charge and critical exponents
for a variety of random circuit models of measurement-induced criticality. We found strong
evidence that the transitions in the Haar, Clifford, and percolation  problems 
belong to three distinct universality classes.
Using the dual unitary variation of these models, we extracted accurate
values for the aforementioned quantities. Additionally,
we have clear evidence of  multifractal
scaling and thus a continuous spectrum of scaling dimensions at the transition in the Haar model.  

\emph{Acknowledgments.}
A.Z. and J.H.P.\ are partially supported by Grant No. 2018058 from the United States-Israel Binational Science Foundation (BSF), Jerusalem, Israel. 
J.H.P.\ acknowledges support from the Alfred P. Sloan Foundation through a Sloan Research Fellowship.
A.Z. is partially supported through a Fellowship from the Rutgers Discovery Informatics Institute. J.H.W. acknowledges the Aspen Center for Physics where part of this work was performed, which is supported by National Science Foundation grant PHY-1607611. R.V. acknowledges support from the Air Force Office of Scientific Research under Grant No. FA9550-21-1-0123, and the Alfred P. Sloan Foundation through a Sloan Research Fellowship. S.G. acknowledges support from NSF DMR-1653271. The authors acknowledge the Beowulf cluster at the Department of Physics and Astronomy of Rutgers University and the Office of Advanced Research Computing (OARC) at Rutgers, The State University of New Jersey (http://oarc.rutgers.edu) for providing access to the Amarel cluster, and associated research computing resources that have contributed to the results reported here.
The Flatiron Institute is a division of the Simons Foundation.
D. A. H. is supported in part by the DARPA DRINQS program.

\bibliographystyle{apsrev4-2}
\bibliography{references}

\widetext
\clearpage
\textbf{\large Supplemental Material: Operator scaling dimensions and multifractality at measurement-induced transitions}

\setcounter{equation}{0}
\setcounter{figure}{0}
\setcounter{table}{0}
\setcounter{page}{1}
\renewcommand{\theequation}{S\arabic{equation}}
\setcounter{figure}{0}
\renewcommand{\thefigure}{S\arabic{figure}}
\renewcommand{\thepage}{S\arabic{page}}
\renewcommand{\thesection}{S\arabic{section}}
\renewcommand{\thetable}{S\arabic{table}}
\makeatletter

\section{\label{sec:equil}Equilibration time}
In this section, we describe our numerical method of computing the free-energy of the implicit statistical mechanics model describing MIPTs in the generic models.  We introduce a method to cleanly distinguish bulk and boundary contributions to the free-energy.

The entropy of the measurement record $F$  can be viewed as an
average of the logarithm of the probability of a given trajectory, i.e.,
$F = -\sum_{\mathbf{m}} p_{\mathbf{m}} \ln p_{\mathbf{m}}=-\sum_{\mathbf{m}}\langle \ln p_{\mathbf{m}} \rangle$.
Here, the expectation value is taken over an ensemble where each trajectory is weighted by its Born probability.
Since the probability of
a given trajectory depends on the product of the Born probabilities of all the
measurements we can write
\begin{equation}
    F = -\sum_{i=1}^{N_\mathrm{meas}} \langle \ln p(m_i\rvert m_{i-1},...,m_1)\rangle,
\end{equation}
where $p(m_i\rvert m_{i-1},...,m_1)$ is the conditional probability of the set
of measurement outcomes $m_i$ given the previous series of measurement
outcomes. This result shows that we can perform a Monte Carlo sampling of the
Born probabilities obtained during the simulation to compute 
$F$.
The entropy is quite sensitive to the initial conditions at early times as we now describe.

\begin{figure}
    \centering
    \subfloat[]{\includegraphics[width=0.48\linewidth]{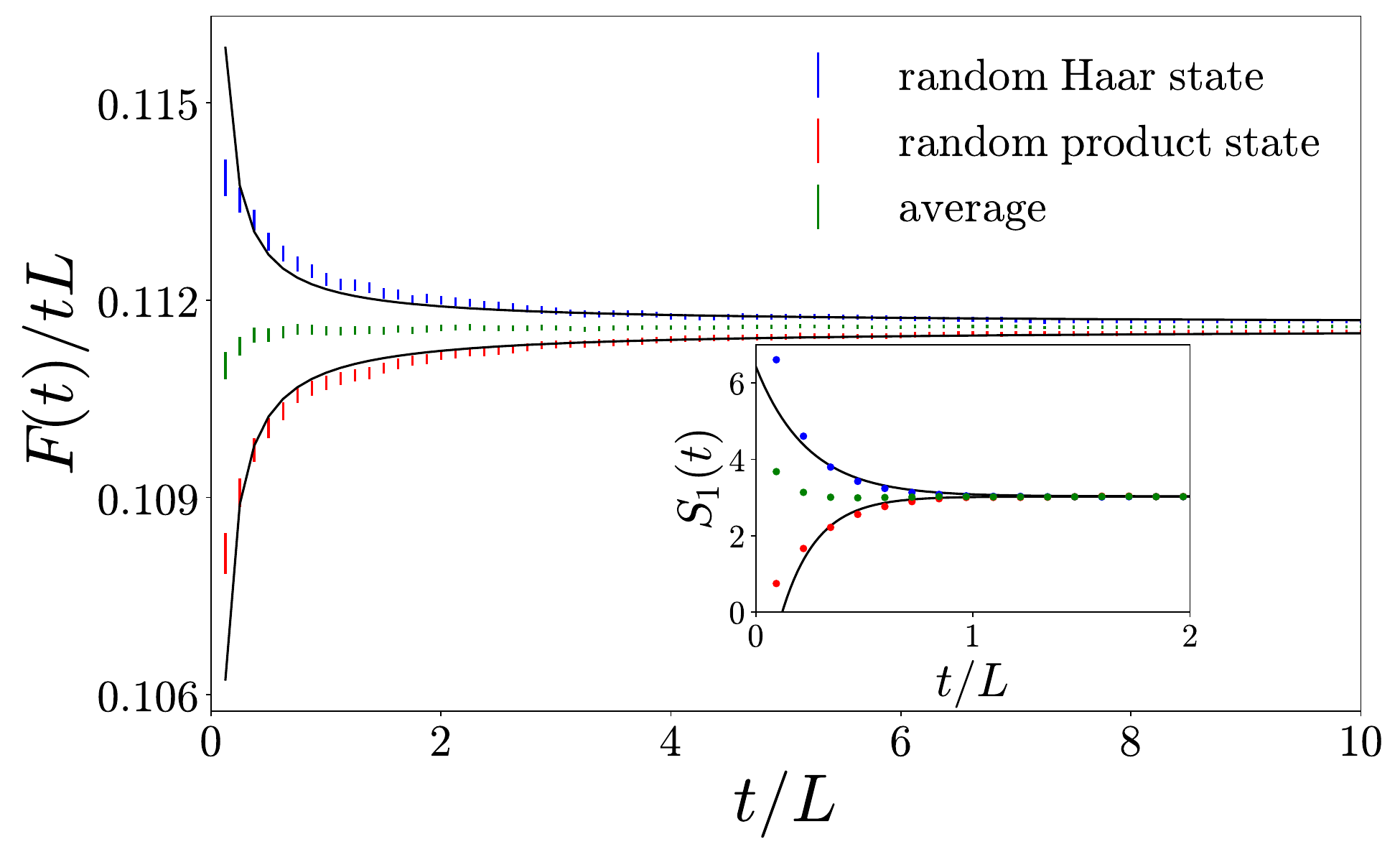}}
    \subfloat[]{\includegraphics[width=0.48\linewidth]{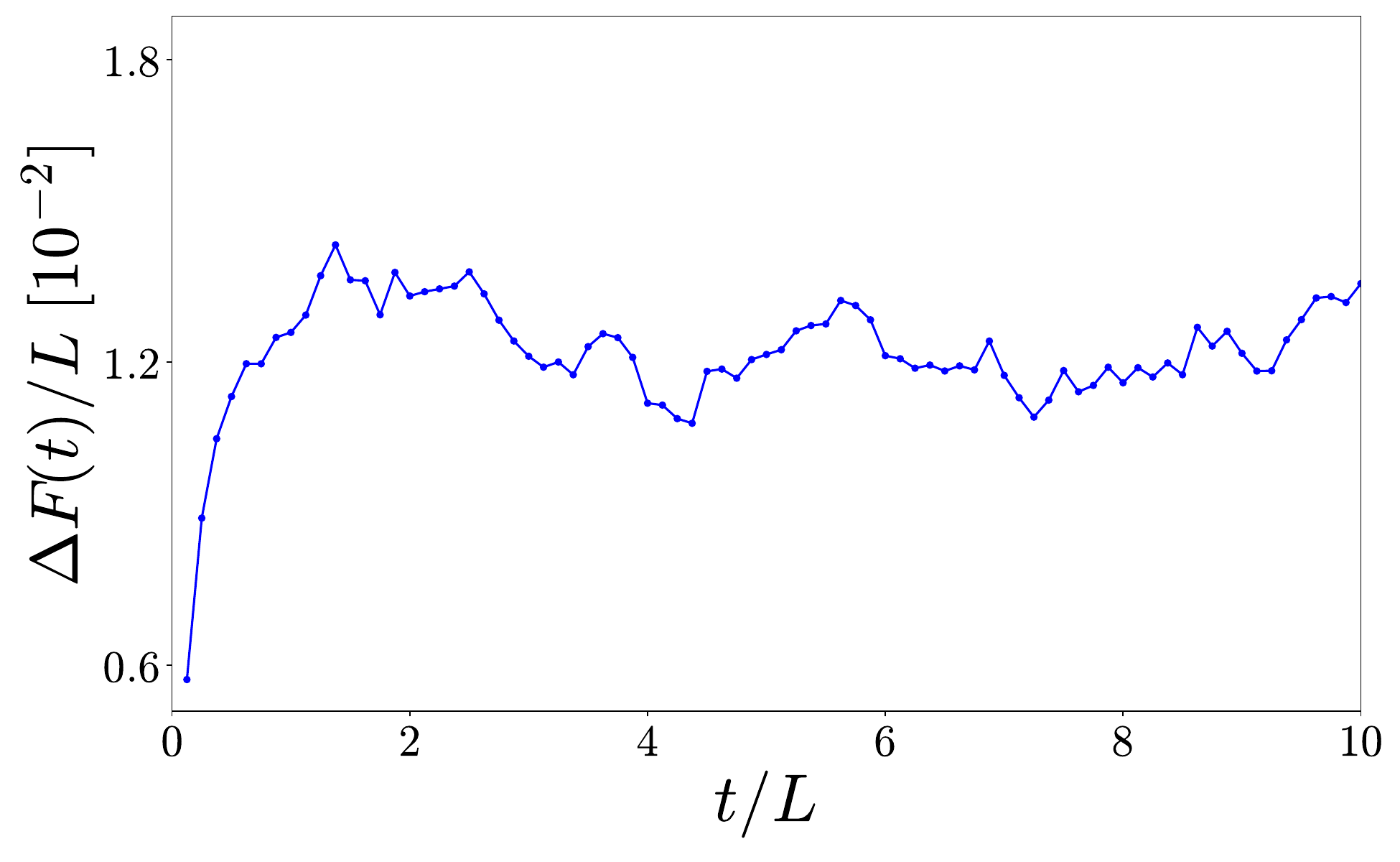}}
    \caption{\label{fig:emr_hp}(a) Average entropy density of the measurement record after a time $t$.
    Due to boundary effects, the entropy density of the measurement
    record averaged separately over random Haar and random product initial
    states saturates slowly to their common asymptotic value. To limit these effects, we wait an
    equilibration time $\tau=4L$ before recording the entropy of the
    measurement record and then average the two results together. Data is
    shown for $L=16$, $p = 0.17$ and $25\,000$ samples. In the inset, we show
    that this saturation criteria agrees well with the half-cut von Neumann entanglement entropy,
    $S_1(t)$.
    (b) The difference between the entropy of the measurement records
    $\Delta F(t) = F^\mathrm{Haar}(t) - F^\mathrm{product}(t)$.
    } 
\end{figure}

To compute the entropy density, we record the entropy
accumulated as a function of time, $F(t)$, and obtain the density from
the slope of the linear fit of the infinite time limit entropy
density of the measurement record $F(t\to\infty)/L$ vs $t$.
As an integral, this quantity, at late times takes the form $F(t\to\infty)/L = F^\mathrm{bdry} + t F^\mathrm{bulk}$ where $F^\mathrm{bdry}$ comes from the choice of initial state and $F^\mathrm{bulk}$ comes from the steady state wave function.
As a result, $F(t\to\infty)/tL = F^\mathrm{bdry}/t + F^\mathrm{bulk}$ indicating a convergence of $1/t$ as indicated in Fig.~\ref{fig:emr_hp}a by the solid lines. 
To uncover when the boundary effects are saturated, we can take two different initial states, a Haar random initial state and a random product state, and compute $\Delta F(t) \equiv F^\mathrm{Haar}(t) - F^\mathrm{product}(t)$ which saturates when the boundary effects saturate; we observe exactly this in Fig.~\ref{fig:emr_hp}b. 
Once saturation is achieved, we can effectively deduce that the wave function has reached the steady state and the average (green) is saturated.
This saturation criteria agrees well with the half-cut entropy shown in the inset of Fig.~\ref{fig:emr_hp}a, and we conservatively obtain $\tau_\mathrm{Haar} \sim \tau_\mathrm{product} \lesssim 2L$ suggesting we
should wait a time $\tau >
2L$ before we begin
recording the entropy of the measurement record.
For our data we have chosen
$\tau=4L$ and recorded the data for an additional time $t_f = 24L$, where one
time step consists of either an even or odd layer of gates and a layer of
measurements. 
To further improve results, after averaging over the random Haar
and product initial states separately, the results are averaged together.
The error in the entropy density of the measurement record is estimated by
computing the entropy density for individual trajectories and performing a
bootstrap analysis~\cite{efron1992bootstrap}. The two initial states are
bootstrapped separately over 1000 samples and their errors are combined using
\begin{equation}
    \sigma = \frac{1}{2}\sqrt{\sigma_\mathrm{Product}^2 + \sigma_\mathrm{Haar}^2}.
\end{equation}

\section{\label{sec:ani}Anisotropy parameter}
In this section, we describe the arguments based on conformal invariance that allow us to efficiently extract the anisotropy parameter at critical points of random circuits with measurements in 1+1 dimensions.

\begin{figure}
    \includegraphics[scale=0.3]{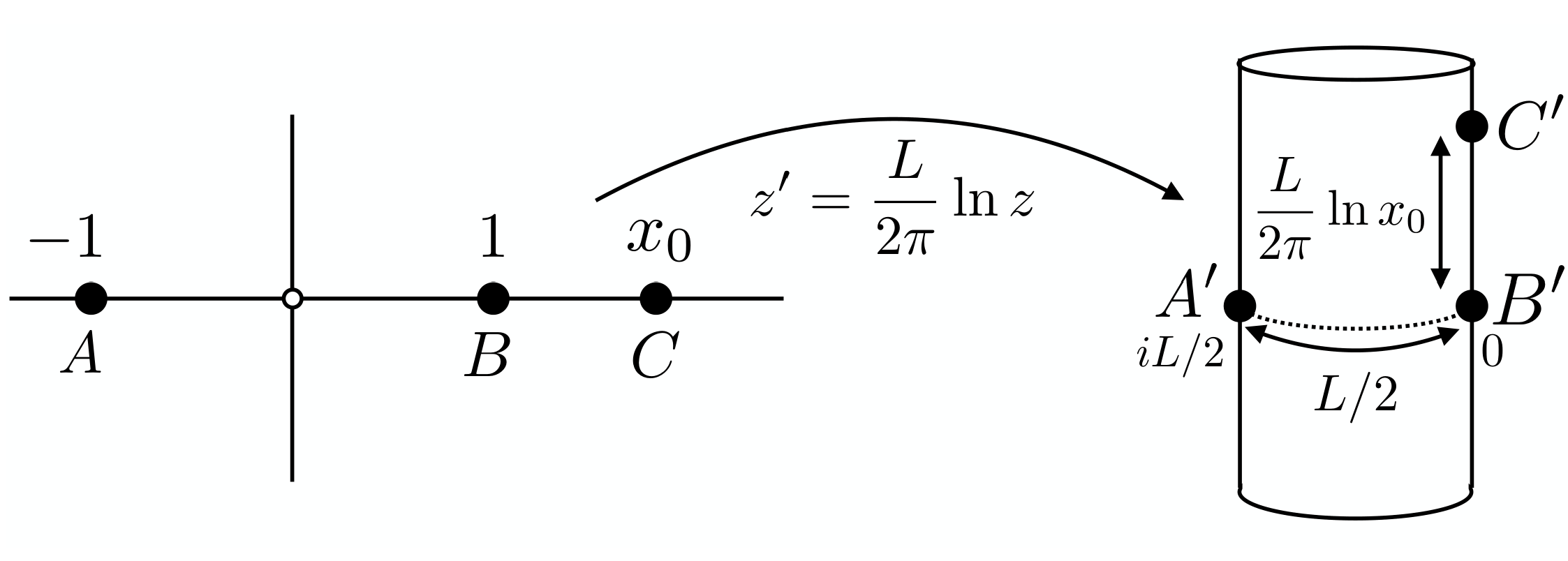}
    \caption{\label{fig:conformal_map}The conformal mapping from a plane to a
    cylinder.}
\end{figure}

To estimate the area, $A=\alpha tL$, that arises in the free energy density, it is necessary to calculate the
anisotropy parameter, $\alpha$, that relates space and time, i.e., $L = \alpha t$.
This parameter can be estimated by comparing the correlation functions along
the space and time directions as we describe below.
Using the conformal mapping $z' = f(z) = \frac{L}{2\pi}\ln z$ (see
Fig.~\ref{fig:conformal_map}), we can relate the correlation functions on the
infinite cylinder, $g'(z_1', z_2')$, to correlation functions on the plane,
$g(z_1,z_2)$, through 
\begin{equation}
    g'(z_1',z_2') = \lvert f'(z_1)\rvert^{-\Delta}\lvert f'(z_2)\rvert^{-\Delta} g(z_1,z_2),
    \label{eq:coor_transf}
\end{equation}
where $\Delta$ is the conformal dimension.
For a 1+1 dimensional CFT,
\begin{equation}
    g(z_1,z_2) = \frac{1}{\lvert z_1 - z_2\rvert^{2\Delta}}
    \label{eq:corr_func_plane}
\end{equation}
and after applying the transformation to Eq.~\eqref{eq:corr_func_plane} we have
\begin{equation}
    g'(z'_1,z'_2) = \left(\frac{\pi}{L}\right)^{2\Delta}\frac{1}{\left\lvert\sinh\left[\frac{\pi}{L}\left(z'_1-z'_2\right)\right]\right\rvert^{2\Delta}}.
\end{equation}
We can extract $\alpha$ from the ratio of the correlation functions
\begin{align}
    g'_\mathrm{space} &= g'(0, iL/2) = \left(\frac{\pi}{L}\right)^{2\Delta} \\
    g'_\mathrm{time} &= g'(0, \alpha t) = \left(\frac{\pi}{L}\right)^{2\Delta}\left(\frac{2e^{\pi\alpha t/L}}{e^{2\pi \alpha t/L} - 1}\right)^{2\Delta} \\
    \frac{g'_\mathrm{time}}{g'_\mathrm{space}} &= \left(\frac{2e^{\pi\alpha t/L}}{e^{2\pi \alpha t/L} - 1}\right)^{2\Delta}.
    \label{eq:corr_func_spacetime_cyl}
\end{align}
To eliminate the dependence on $\Delta$, we look for the matching time, $t_*$,
at which the space and time correlation functions acquire the same value.
Setting $g'_\mathrm{time}/g'_\mathrm{space} = 1$ in
Eq.~\eqref{eq:corr_func_spacetime_cyl}, the resulting quadratic equation can be
solved for the anisotropy parameter
\begin{equation}
    \begin{split}
    e^{2\pi\alpha t_*/L} - 2 e^{\pi\alpha t_*/L} - 1 = 0 \\
    \implies \alpha = \ln\left(1+\sqrt{2}\right)\frac{L}{\pi t_*}.
    \end{split}
\end{equation}

To compute this numerically, we calculate the mutual information between two
initially locally entangled reference qubits. We run the unitary-measurement
dynamics out to $\tau_1 = 4L$, measure site $x_1$ and entangle this site with a
reference qubit. We then run the dynamics out to $\tau_2$, measure site $x_2$
and entangle this site with another reference qubit.  After this second event,
we follow the mutual information $I_{12}(x_1, x_2, \tau_1, \tau_2)$ between the
two reference qubits as a probe of the order parameter correlations.
We use a space like separation of $\delta x = \lvert x_2 - x_1\rvert = L/2$
with $\delta \tau = 0$ to determine $g'_\mathrm{space}$ and time like separation
of $\delta \tau = \tau_2 - \tau_1$ with $\delta x = 0$ to determine $g'_\mathrm{time}$.

\section{Lyapunov exponents}

In this section, we describe a procedure that only
requires storing a set of vectors that are iterated upon in order to compute
the Lyapunov exopnents.

The Lyapunov exponents of the transfer matrix can be related to the free energy densities
that are used in the calculation of the scaling dimensions of operators in the theory.
However, working with the full transfer matrix becomes exponentially difficult
in the system size and an alternative approach is needed. 

\begin{figure}
    \centering
    \subfloat[]{\includegraphics[width=0.33\linewidth]{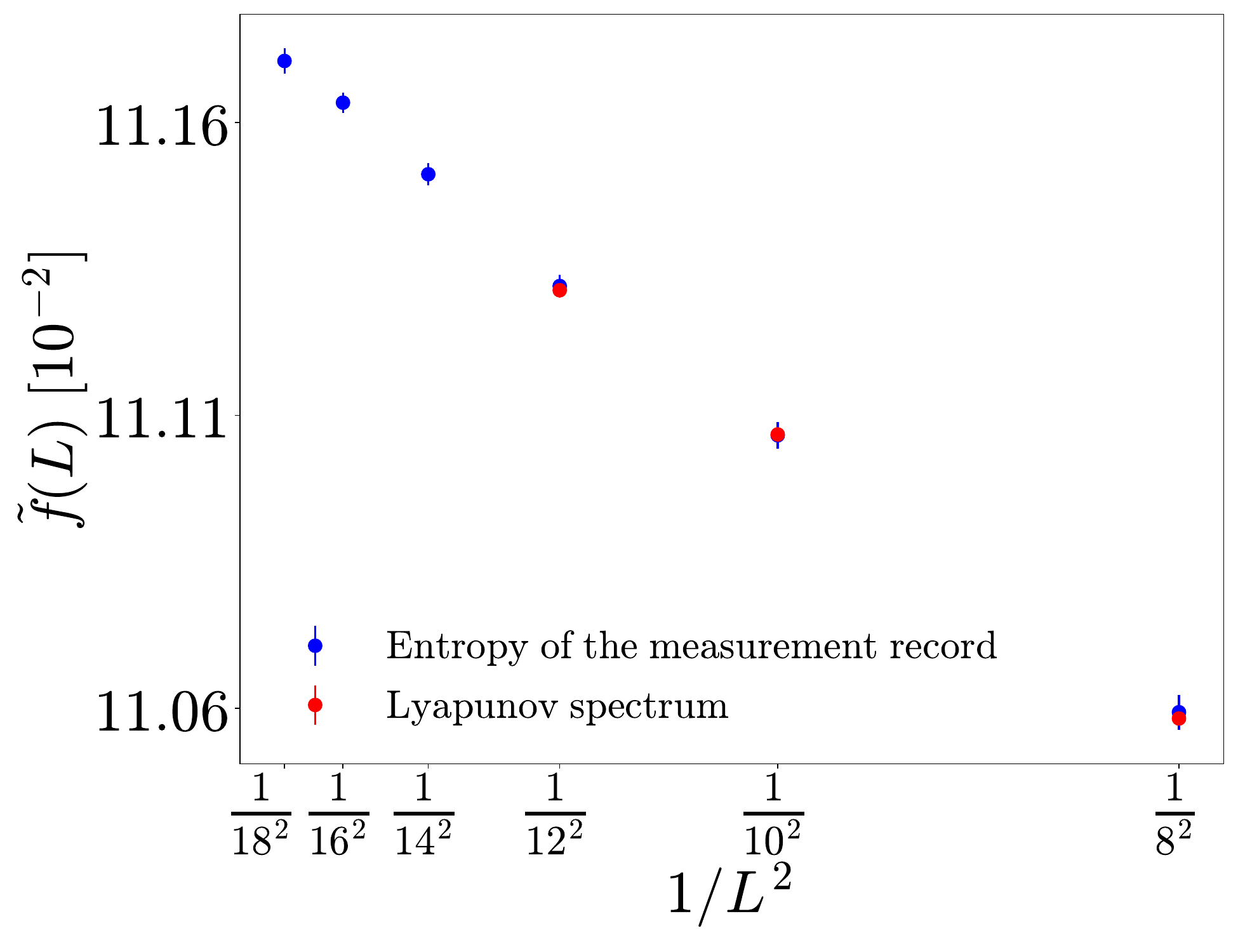}}
    \subfloat[]{\includegraphics[width=0.33\linewidth]{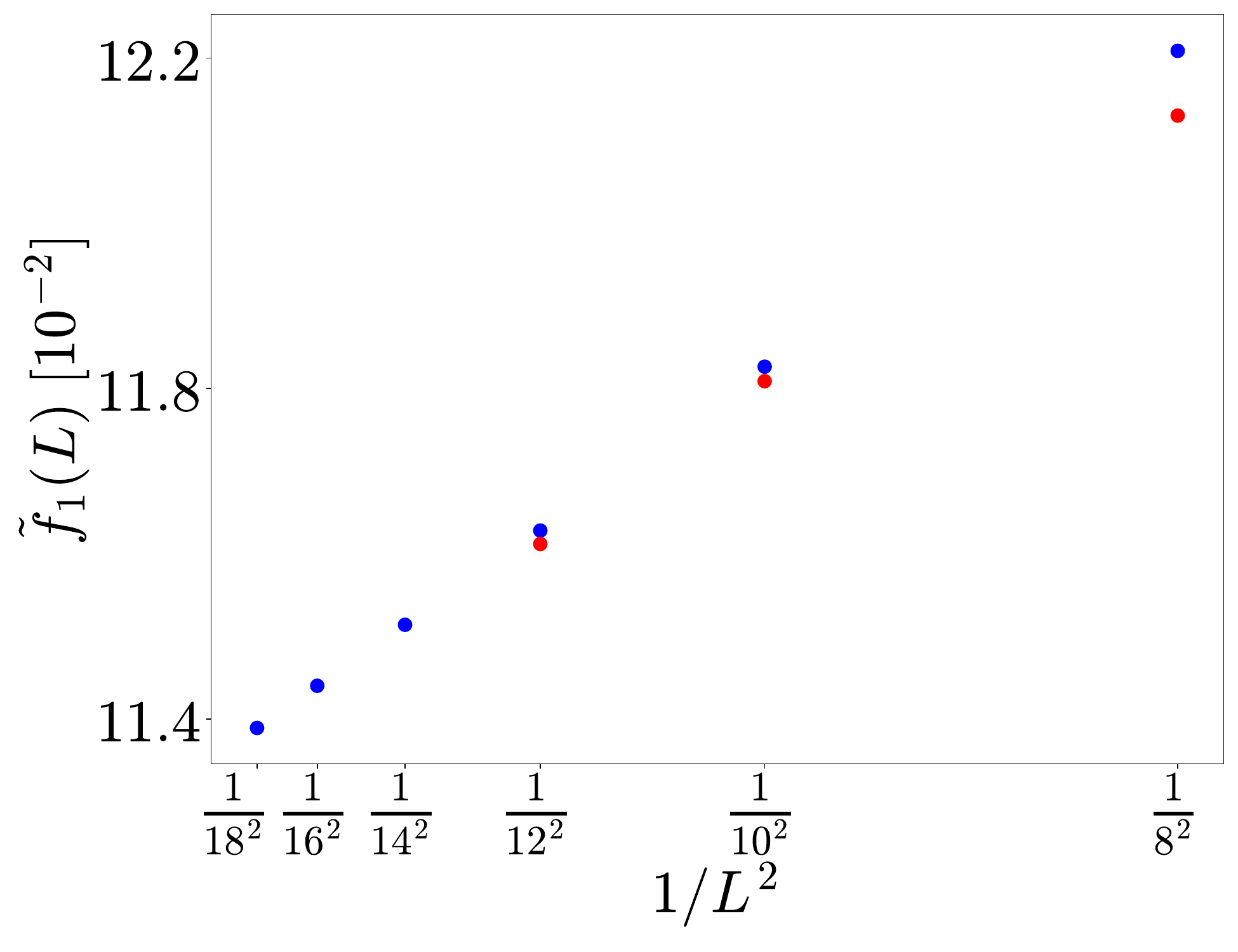}}
    \subfloat[]{\includegraphics[width=0.33\linewidth]{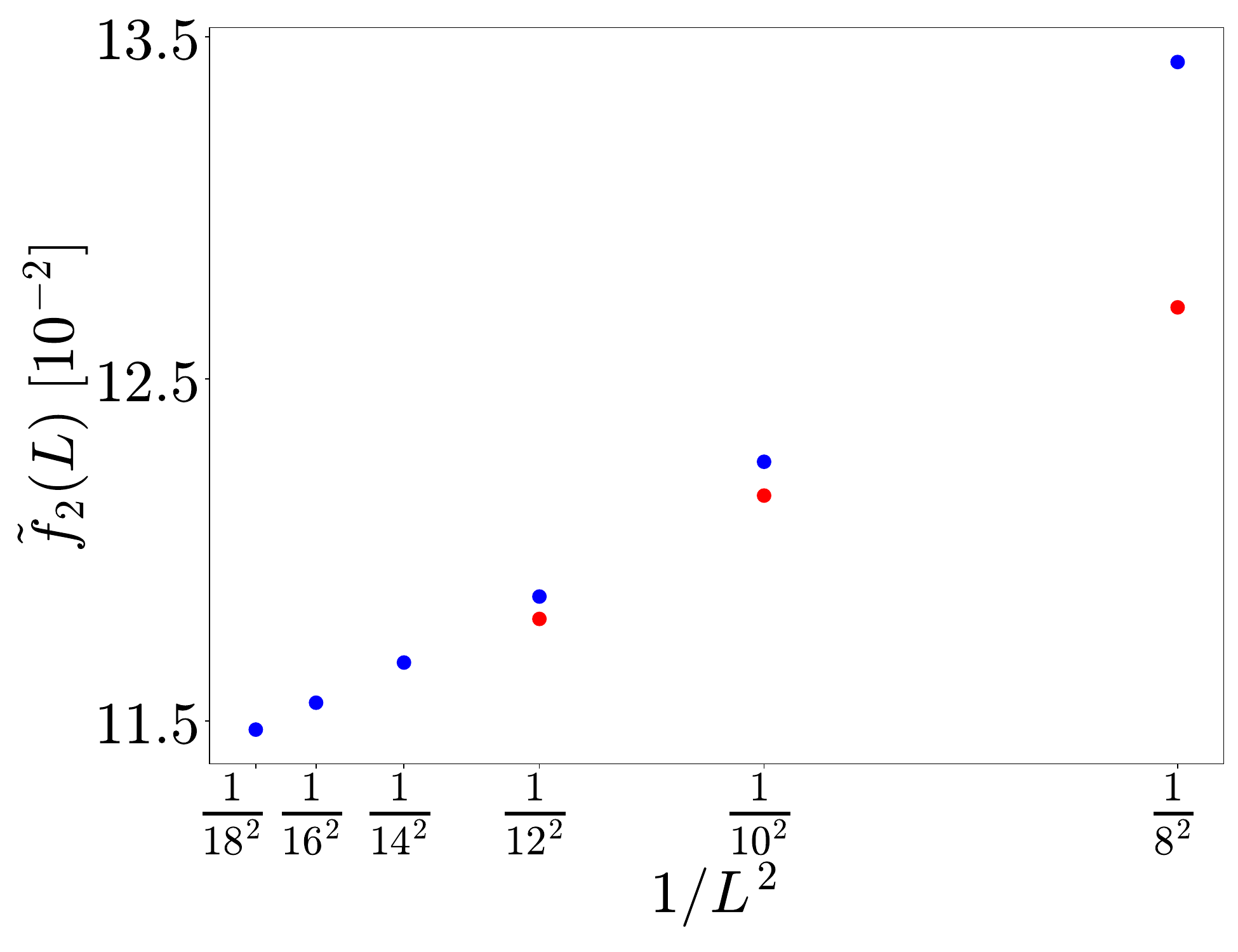}}
    \caption{\label{fig:tm_vec}Comparison of the free energy density obtained using
    the entropy of the measurement record based on the orthogonal vectors with the 
    free energy density obtained using the Lyapunov spectrum based on the singular values
    of the transfer matrix. At small system sizes, the Gram-Schmidt orthogonaliztion
    procedure quickly zeros out the higher order orthogonal vectors making it difficult to
    sample at late times. The tilde in the free energy density denotes that $\alpha$ is not taken
    into account in the area.
    }
\end{figure}

We are interested in characterizing the large $m$ behavior of the application
of i.i.d. random transfer matrices $\mathcal{T}_j \in \mathbb{C}^{d\times d}$ to a vector
$\ket{v_0} \in \mathbb{C}^d$ with $j = 1,2,\ldots,m$. This evolution can be
described by the recurrence
\begin{equation}
    \ket{v_0^{(j)}} = \mathcal{T}_j\ket{v_0^{(j-1)}}, \qquad j = 1,2,...,m
\end{equation}
for some initial normalized vector $\ket{v_0}$. The randomness of the matrices
$\mathcal{T}_j$ implies the choice of the probabily measure on $\mathbb{C}^{d\times d}$.
The large $m$ behavior can be characterized by considering the leading Lyapunov
exponent found by the Furstenberg method
\begin{equation}
    \lambda_0(L) = \lim_{m\rightarrow\infty} \frac{1}{m} \mathbb{E} \log \left\Vert \ket{v^{(m)}_0} \right\Vert
    \label{eq:lyap_exp}
\end{equation}
where $\mathbb{E}$ denotes the expectation over the random matrices.
Equation~\eqref{eq:lyap_exp} is independent of the initial vector $\ket{v_0}$
for almost all realizations of the matrices $\mathcal{T}_j$. An alternative definition
that makes the independence on $\ket{v_0}$ explicit is
\begin{equation}
    \lambda_0(L) = \lim_{m\rightarrow\infty} \frac{1}{m} \mathbb{E} \log \left\Vert \prod_{j=1}^m \mathcal{T}_j \right\Vert
\end{equation}
where the matrix norm is the 2-norm, so that $\left\Vert \prod_{j=1}^m \mathcal{T}_j
\right\Vert$ is the largest singular value of $\prod_{j=1}^m \mathcal{T}_j$. The average
free energy per site is related to the leading Lyapunov exponent by
\begin{equation}
    f(L) = -\frac{1}{\alpha L}\lambda_0(L).
\end{equation}
Similarly, the generalized free energies can be related to the higher order Lyapunov
exponents through $f_i(L) = -\frac{1}{\alpha L}\lambda_i(L)$. In order to numerically
compute $\lambda_i$, we can consider of a set of $n$ orthogonal vectors
$\left\{\ket{v_k}\right\}, k = 0, 1, \ldots, n-1$ and iteratively apply the
transfer matrices, $\mathcal{T}_j$. After each application of $\mathcal{T}_j$, the set must be
orthogonalized again.
In Fig.~\ref{fig:tm_vec} we show that the value of the free energy density
obtained from the Lyapunov spectrum approaches that from the entropy of the
measurement record. At small system sizes, the Gram-Schmidt orthogonalization
procedure quickly zeros out the orthogonal vectors making it difficult to
sample at late times. Note that for the vectors $k > 0$, the entropy of the measurement
record, $F_k$, must be slightly modified to account for the orthogonalization procedure and is
given by
\begin{equation}
    F_{k} = -\sum_{j=1}^{m} \ln\norm{P_{\mathrm{GS},k}^{(j)}P_\mathrm{M}^{(j)}\ket{v_k^{(j-1)}}}^2,
\end{equation}
where $P_{\mathrm{GS},k}^{(j)}$ is a projector from the Gram-Schmidt process
and $P_\mathrm{M}^{(j)}$ is a projector onto the meaurement outcomes in $\mathcal{T}_j$.

\section{\label{sec:haar}Haar random circuit}
In this section, we estimate the anisotropy parameter for the Haar random circuit
and use it to compute the effective central charge and scaling dimensions of operators
in theory. We also show evidence of multifractal scaling at the critical point.

\begin{figure}
    \centering
    \subfloat[\label{fig:haar_ani_far}]{\includegraphics[width=0.5\linewidth]{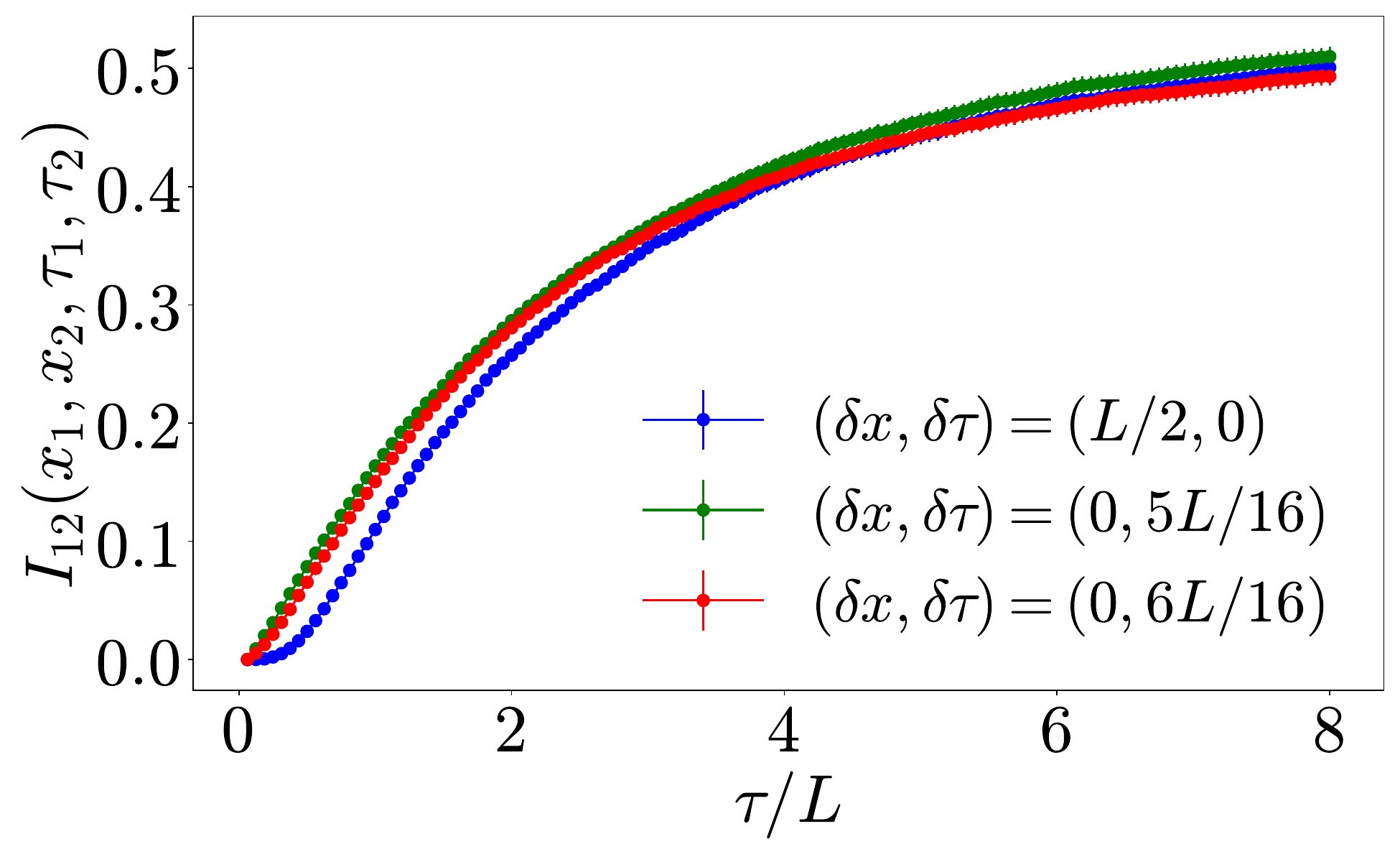}}
    \subfloat[\label{fig:haar_ani_close}]{\includegraphics[width=0.5\linewidth]{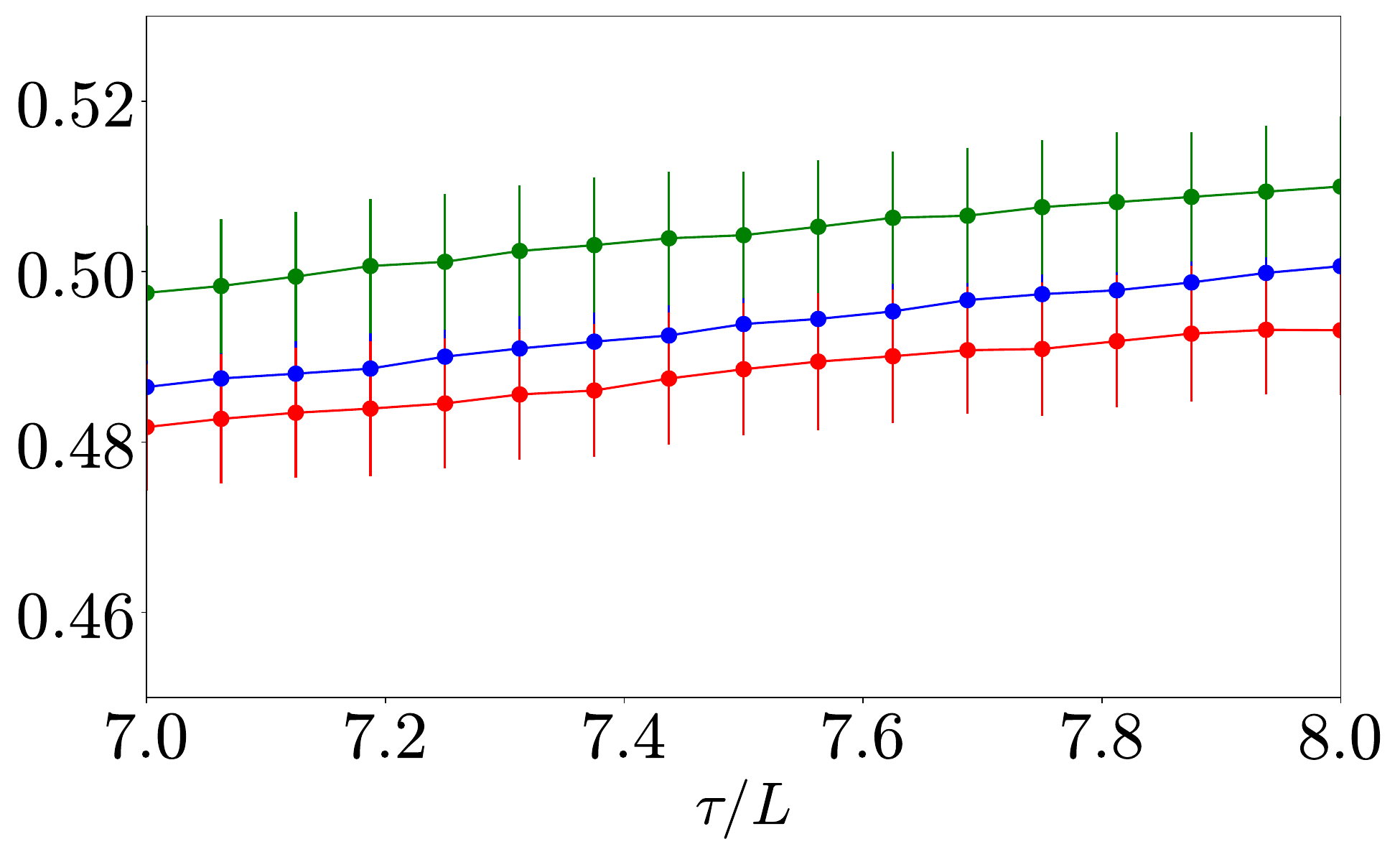}}
    \caption{\label{fig:haar_ani}Haar model space and time correlation functions at $p_c = 0.170$ for $L=16$.
    The matching time, $t_*$, lies between $t=5L/16$ and $t=6L/16$. Using a linear interpolation we
    estimate the true matching time to be $t_* = 5.55$ and, therefore, $\alpha = 0.81(9)$.}
\end{figure}

We can compute the anisotropy parameter for the Haar random circuit using the
procedure described in Sec.~\ref{sec:ani}. The correlation functions along
the space and time directions are shown in Fig.~\ref{fig:haar_ani}. Numerically
computing the correlation functions shows that the matching time is between $t
= 5L/16$ and $t=6L/16$. Performing a linear interpolation
\begin{equation}
    t_* = t_5 + \left[ I_\mathrm{space} - I(t_5)\right] \frac{t_6-t_5}{I(t_6)-I(t_5)}
\end{equation}
which gives $t_* = 5.55$ and $\alpha = 0.81 \pm 0.09$ with the error bar
spanning the range $t_* \in \left[ t_5, t_6\right]$.

\begin{figure}
    \centering
    \includegraphics[width=0.5\linewidth]{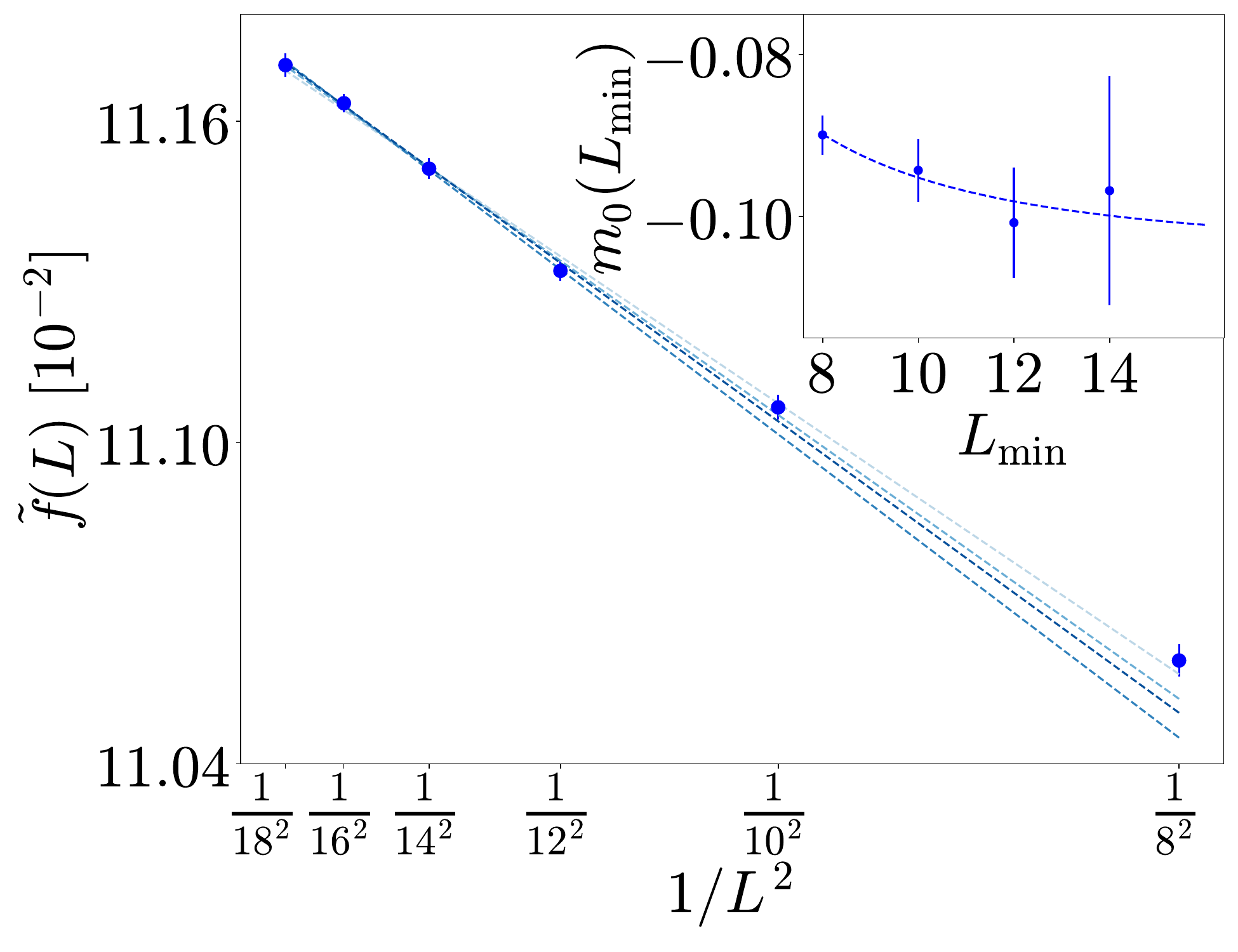}
    \caption{\label{fig:haar_ceff}The free energy density of the Haar model shows
    the expected $1/L^2$ scaling.
    Using the double fitting procedure described in the main text we can extract
    $c_\mathrm{eff}$ from the slope of the free energy density.
    For the Haar model we find $c_\mathrm{eff} = 0.25(3)$, in agreement with the dual unitary result
    placing the two models into the same universality class. The tilde in the free energy density, $\tilde{f}$, indicates
    that $\alpha$ has not been taken into account in the area. Darker blue indicates larger values of
    $L_\mathrm{min} = 8\to 14$.}
\end{figure}

\begin{figure}
    \centering
    \subfloat[\label{fig:f1f0}]{\includegraphics[width=0.32\linewidth]{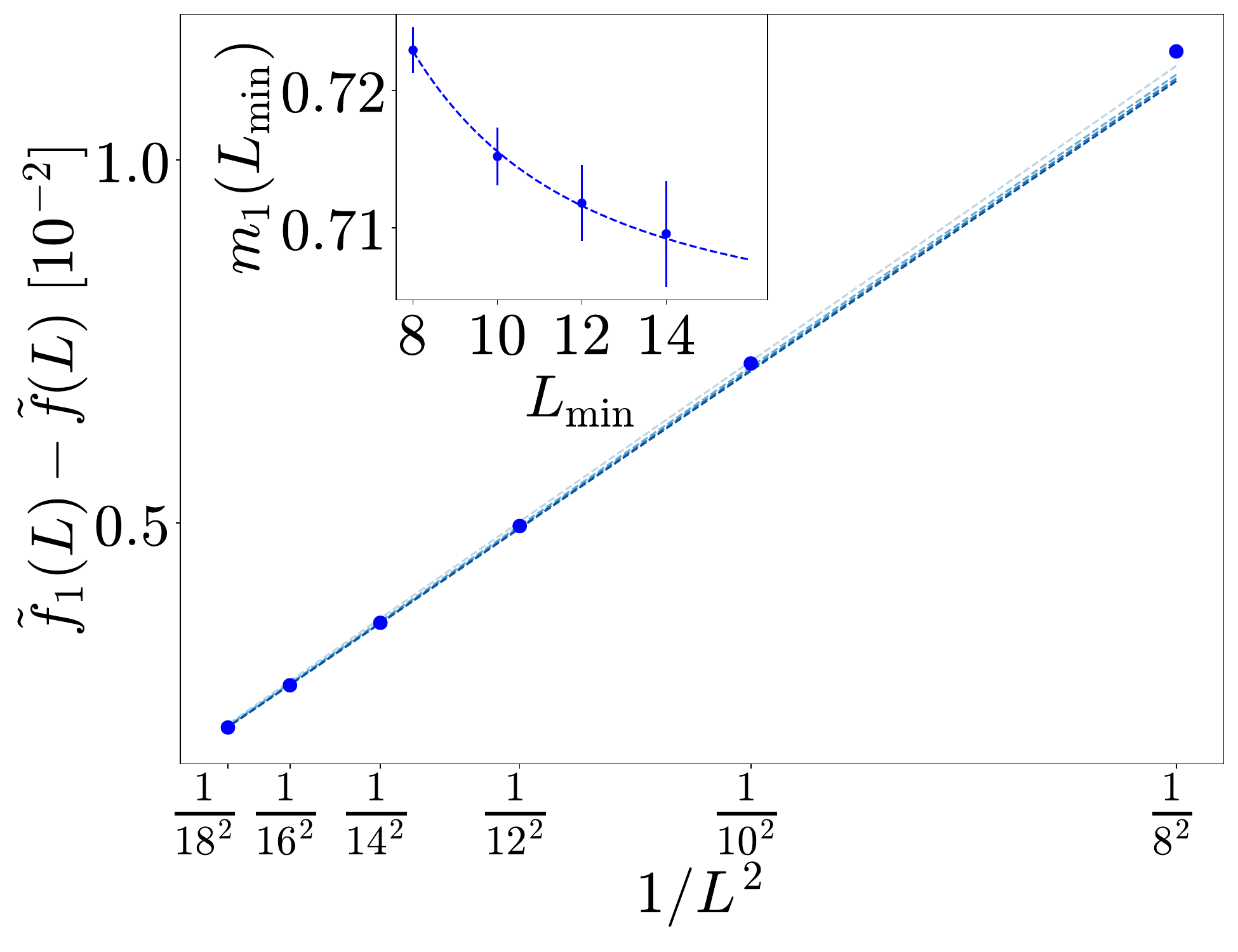}}
    \subfloat[\label{fig:f2f0}]{\includegraphics[width=0.32\linewidth]{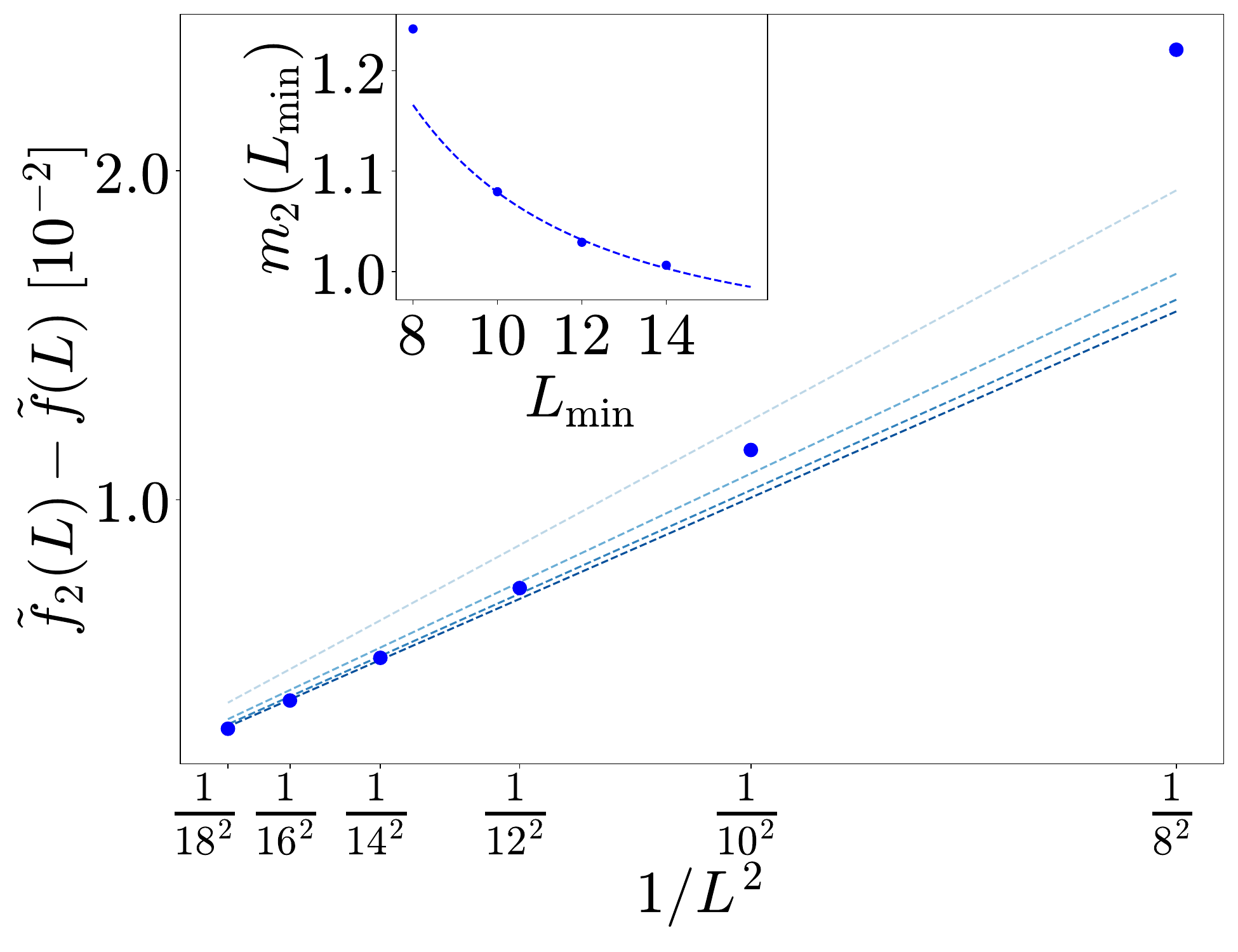}}
    \subfloat[\label{fig:f3f0}]{\includegraphics[width=0.32\linewidth]{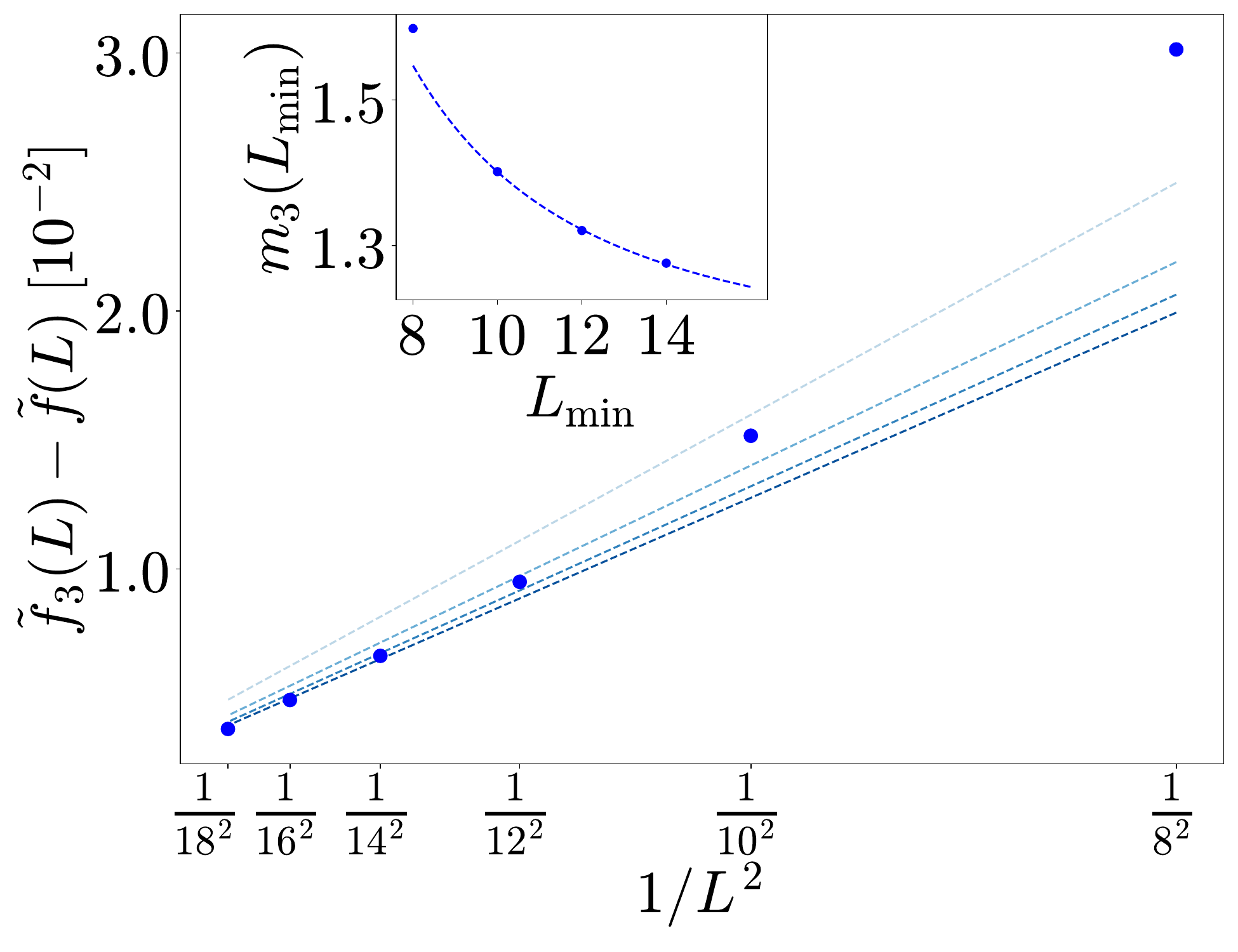}}
    \caption{\label{fig:haar_fif0}
    Scaling of the generalized free energies in the Haar model.
    The differences between the free energy densities
    can be related to the scaling dimension, $x_i^\mathrm{typ}$, of operators in the theory. In the figure, the tilde, e.g. $\tilde{f}_i$,
    denotes that we have no included the anisotropy parameter into the area and it will be introduced in the
    final result. Using a similar double fitting procedure as for the effective central charge, the typical
    values of the first three scaling dimensions are estimated to be
    $x_1^\mathrm{typ} = 0.14(2), x_2^\mathrm{typ} = 0.18(2), x_3^\mathrm{typ} = 0.23(3)$. Darker blue indicates larger
    values of $L_\mathrm{min} = 8\to 14$.}
\end{figure}

\begin{figure}
    \centering
    \includegraphics[width=0.5\linewidth]{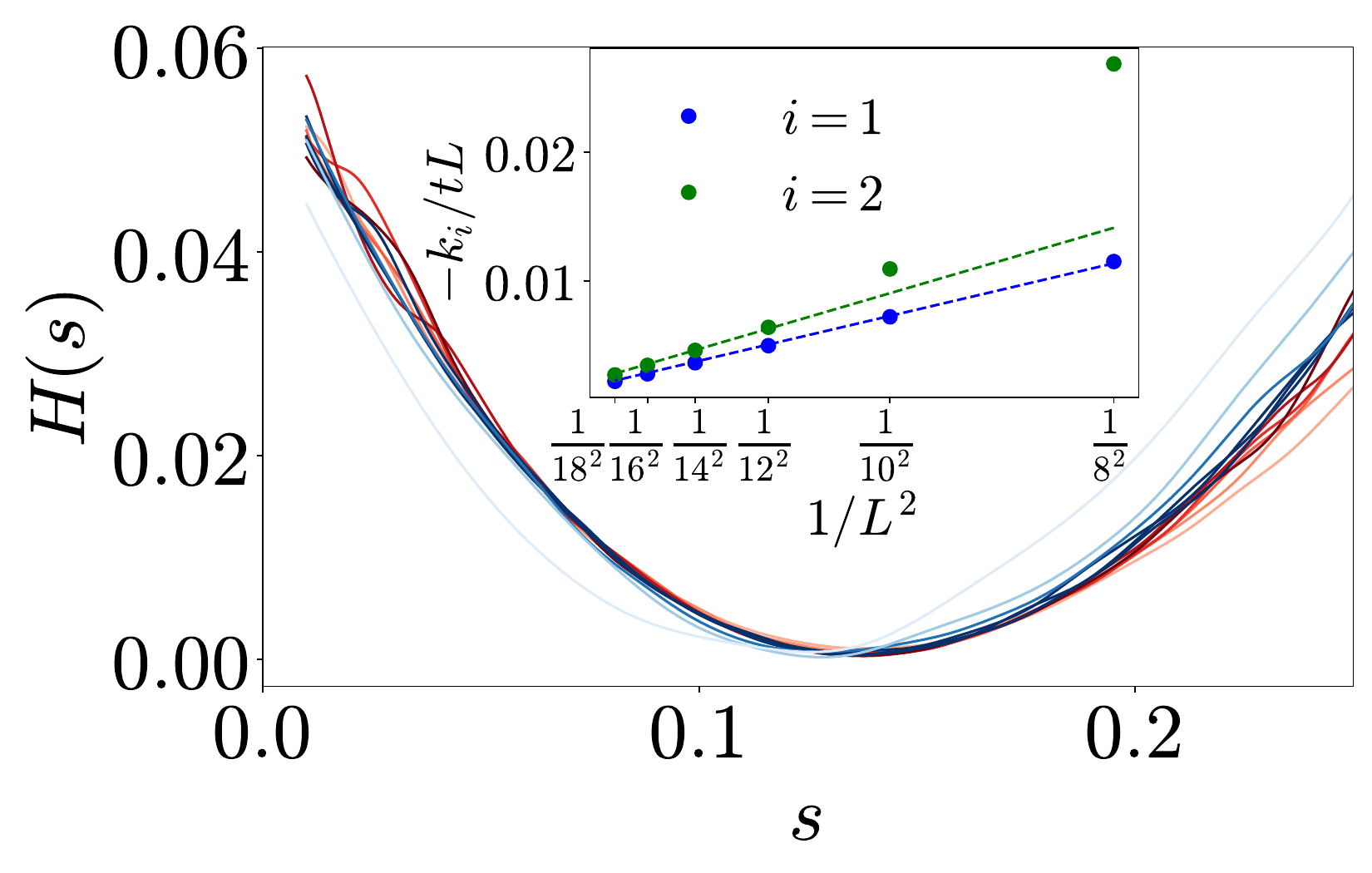}
    \caption{\label{fig:haar_halpha}
    The scaling collapse of the data onto a universal multifractal scaling
    function $H(s)$ demonstrates multifractality at the
    critical point of the Haar transition and corresponds to a continuum of critical exponents.
    Data is shown for the Haar model where darker red indicates larger system sizes
    ($L = 8\to 18, t = 24 L$) and darker blue indicates later times ($t = 3L \to 24L, L = 16$).
    (inset) The first two cumulants $k_i$ of $\ln G_1(t)$
    divided by $tL$ show the expected $1/L^2$ behavior.
    }
\end{figure}

This anisotropy parameter can be incorporated with the results of the free energy density scaling shown
in Fig.~\ref{fig:haar_ceff} to estimate $c_\mathrm{eff} = 0.25(3)$. Note that we have introduced a tilde
into the notation of the free energy density, $\tilde{f}$, to indicate that it does not contain $\alpha$ in the area. The
fit to the slope of $\tilde{f}(L)$ in the inset is given by $m_0(L) = -0.105 + \frac{0.958}{L^2}$.
We can also compute the critical exponents, $x_i^\mathrm{typ}$, from the differences of the
generalized free energy densities as shown in Fig.~\ref{fig:haar_fif0}. Performing the double fitting
procedure and incorporating $\alpha$ into the result we find
$x_1^\mathrm{typ} = 0.14(2)$, $x_2^\mathrm{typ} = 0.18(2)$, $x_3^\mathrm{typ} = 0.23(3)$. The fits in the inset are given by
$m_1(L) = 0.703 + \frac{1.30}{L^2}$,
$m_2(L) = 0.924 + \frac{15.5}{L^2}$,
and $m_3(L) = 1.14 + \frac{25.9}{L^2}$.
Additionally, we find evidence of multifractality
at the critical point based on the data collapse of $H(s)$ as well as the scaling of the cumulants
of $\ln G_1(t)$, see Fig.~\ref{fig:haar_halpha}.

\section{Dual unitary}
In this section, we determine the critical point of the dual unitary model
using the entanglement transition order parameter. At the critical point,
we verify that $\alpha = 1$
and use it to compute the effective central charge and scaling dimensions of operators
in theory. 

As argued in the main text, the transition in the dual unitary model
lies in the same universality class as that of the generic Haar model and is used
to provide a more accurate estimate of the quantities calculated as it
constrains $\alpha=1$.

\begin{figure}
    \centering
    \subfloat[\label{fig:DU_op}]{\includegraphics[width=0.32\linewidth]{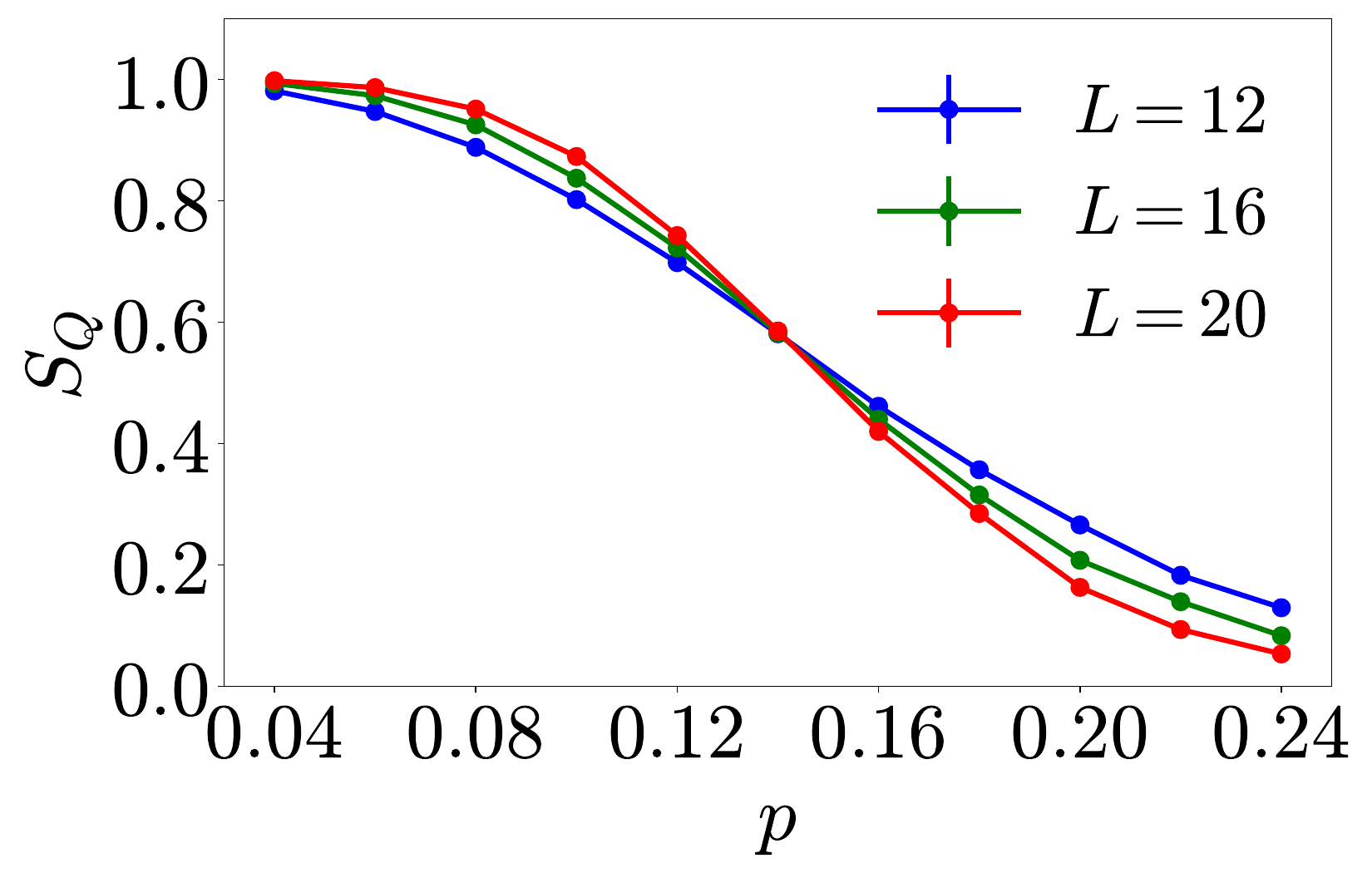}}
    \subfloat[\label{fig:du_ani}]{\includegraphics[width=0.32\linewidth]{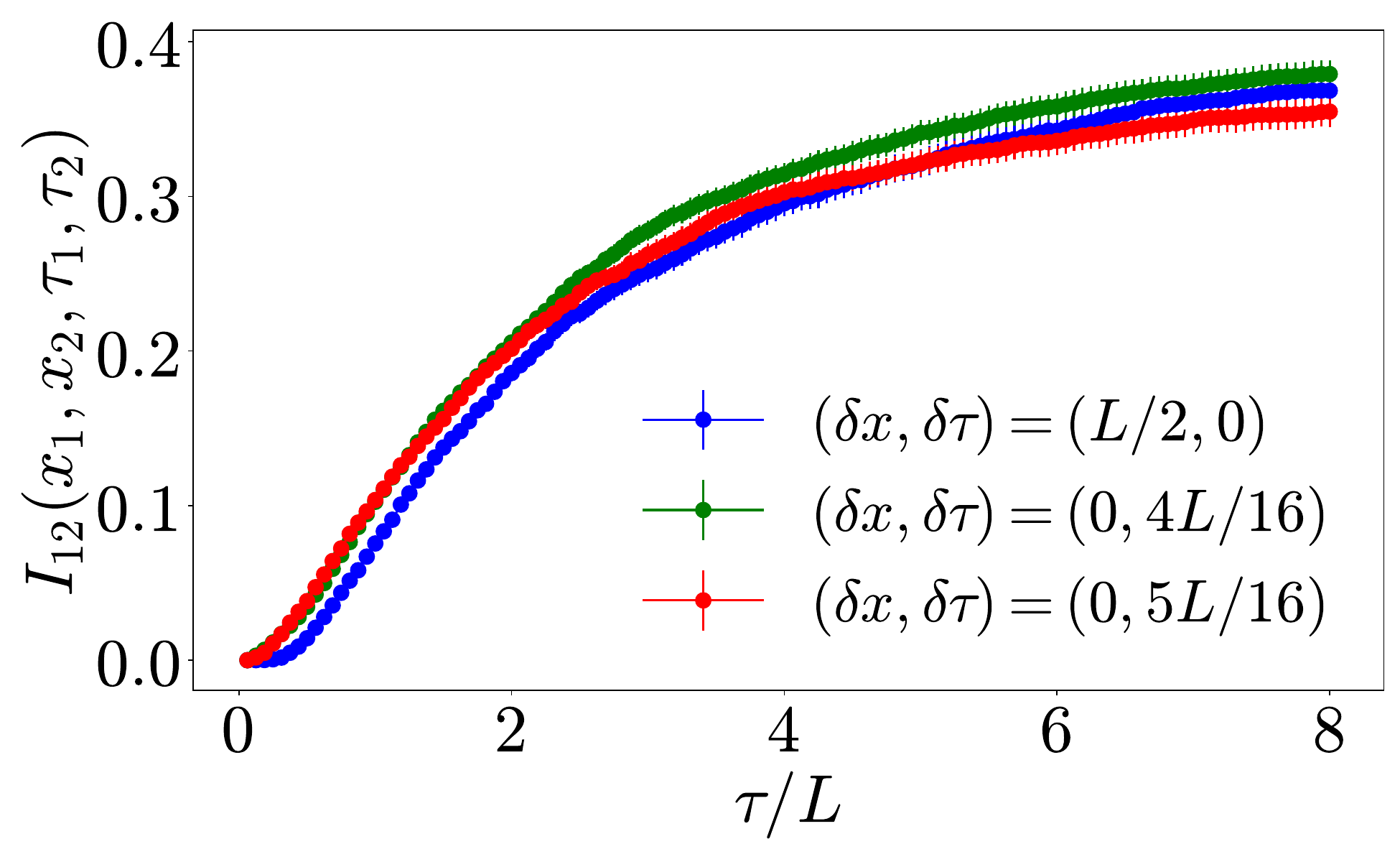}}
    \subfloat[\label{fig:du_ani_close}]{\includegraphics[width=0.32\linewidth]{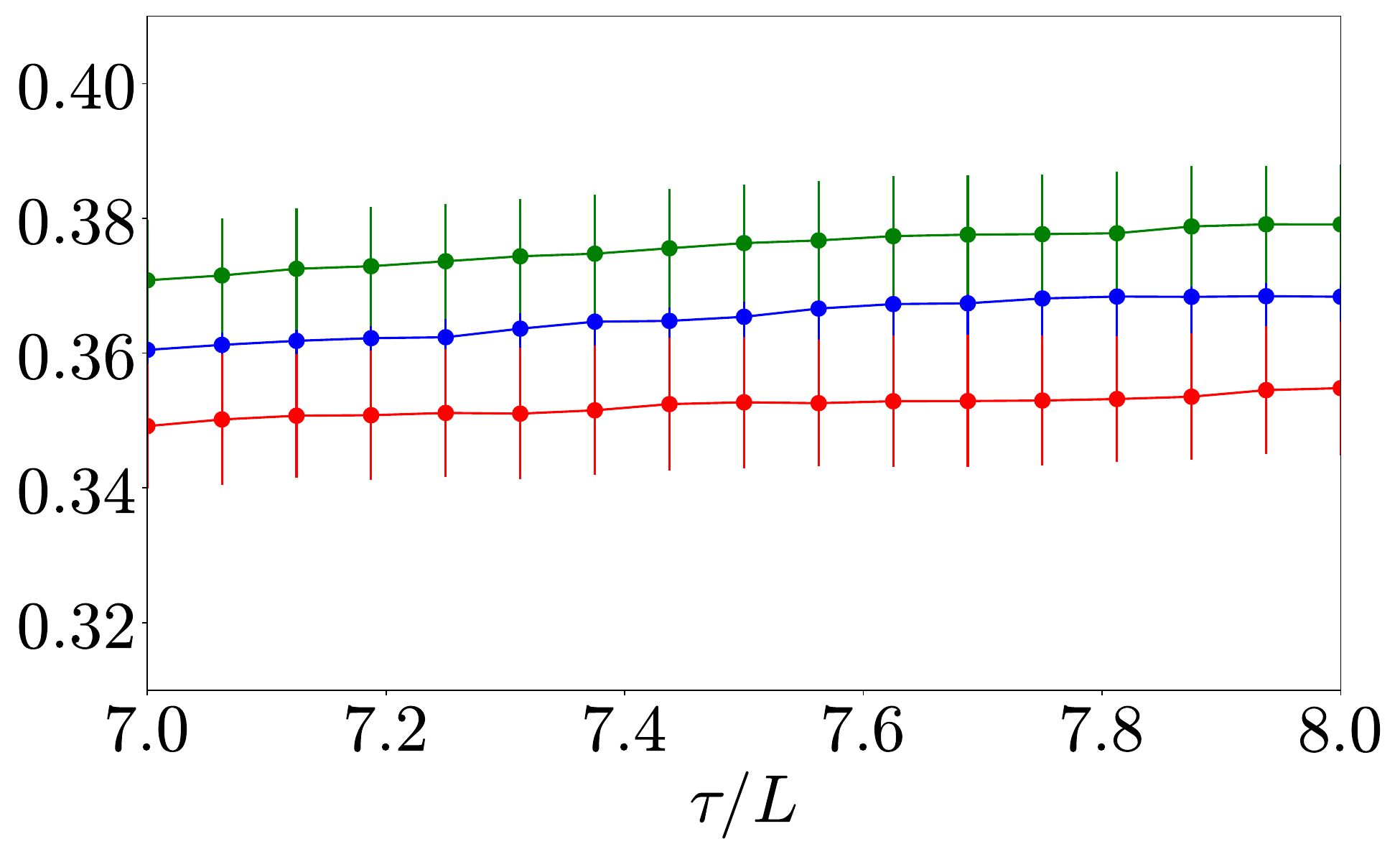}}
    \caption{
    (a) Dual unitary order parameter $S_Q$ as a function of the measurement probability, $p$.
    The critical point is indicated by the crossing at $p_c = 0.14(1)$.
    (b) and (c) Dual unitary space and time correlation functions at $p_c = 0.140$ for $L=16$. The matching time, $t_*$, at which the space and time correlation functions
    coincide lies between $t = 4L/16$ and $t = 5L/16$. Using a linear interpolation we can estimate
    the true matching time to be $t_* = 4.44$ and, therefore, $\alpha = 1.0(1)$ in agreement with the
    expectation that $\alpha = 1$.}
\end{figure}

The dual unitary circuit model we consider consists of 2-qubit gates of the
form~\cite{bertini2019exact}
\begin{equation}
    U = e^{i\phi}(u_+ \otimes u_-) \cdot V[J] \cdot (v_- \otimes v_+)
\end{equation}
where $\phi, J \in \mathbb{R}, u_\pm, v_\pm \in \mathrm{SU}(2)$ and
\begin{equation}
    V[J] = \exp[-i\left(\frac{\pi}{4}\sigma^x \otimes \sigma^x + \frac{\pi}{4}\sigma^y \otimes \sigma^y + J\sigma^z \otimes \sigma^z\right)].
\end{equation}
With this choice, $U$ is unitary in both the space and time directions, i.e.,
$U^\dagger U = \tilde{U}^\dagger\tilde{U} = \mathbb{1}$ where
\begin{equation}
    \bra{k}\otimes\bra{l}\tilde{U}\ket{i}\otimes\ket{j} = \bra{j}\otimes\bra{l}U\ket{i}\otimes\ket{k}.
\end{equation}
In the numerical simulations we choose $\phi, J$ uniformly from $[0, 2\pi)$.
To find the critical point we look at the order parameter as a function of the
measurement probability $p$.  This is the best measure of the critical point
since there is a strong even/odd effect in the tripartite mutual information
($\mathcal{I}_3$) data.
In Fig.~\ref{fig:DU_op} we see a clear crossing of the order parameter at $p_c = 0.14(1)$.
Using this critical point we can estimate the anisotropy parameter $\alpha$
by measuring correlation functions along the space and time dimensions as
described in Sec.~\ref{sec:ani}. Numerically computing the correlation
functions shows that the matching time is between $t = 4L/16$ and $t = 5L/16$,
see Figs.~\ref{fig:du_ani} and~\ref{fig:du_ani_close}. We can get a better estimate of this
time by performing a linear interpolation
\begin{equation}
    t_* = t_4 + \left[ I_\mathrm{space} - I(t_4)\right] \frac{t_5-t_4}{I(t_5)-I(t_4)}
\end{equation}
which gives $t_* = 4.44$ and $\alpha = 1.0 \pm 0.1$ with the error bar
spanning the range $t_* \in \left[ t_4, t_5\right]$. This result is in
agreement with our expectation that $\alpha = 1$ by the construction of the
gates. In what follows, we take $\alpha$ to be exactly one, thereby,
eliminating the parameter from the calculations and reducing the error bars in
the estimates of all quantities for this model.

\begin{figure}
    \centering
    \subfloat[\label{fig:duf1f0}]{\includegraphics[width=0.32\linewidth]{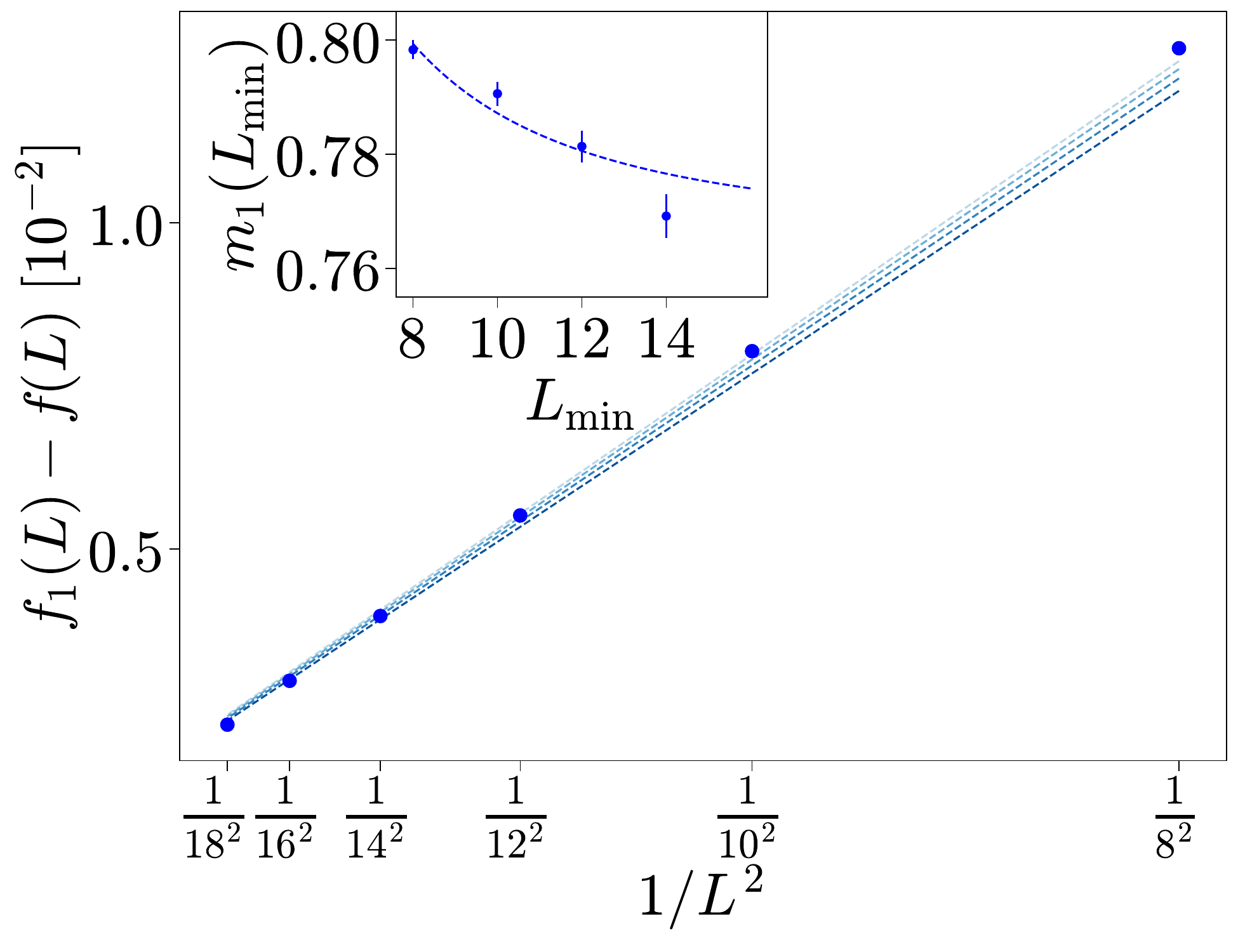}}
    \subfloat[\label{fig:duf2f0}]{\includegraphics[width=0.32\linewidth]{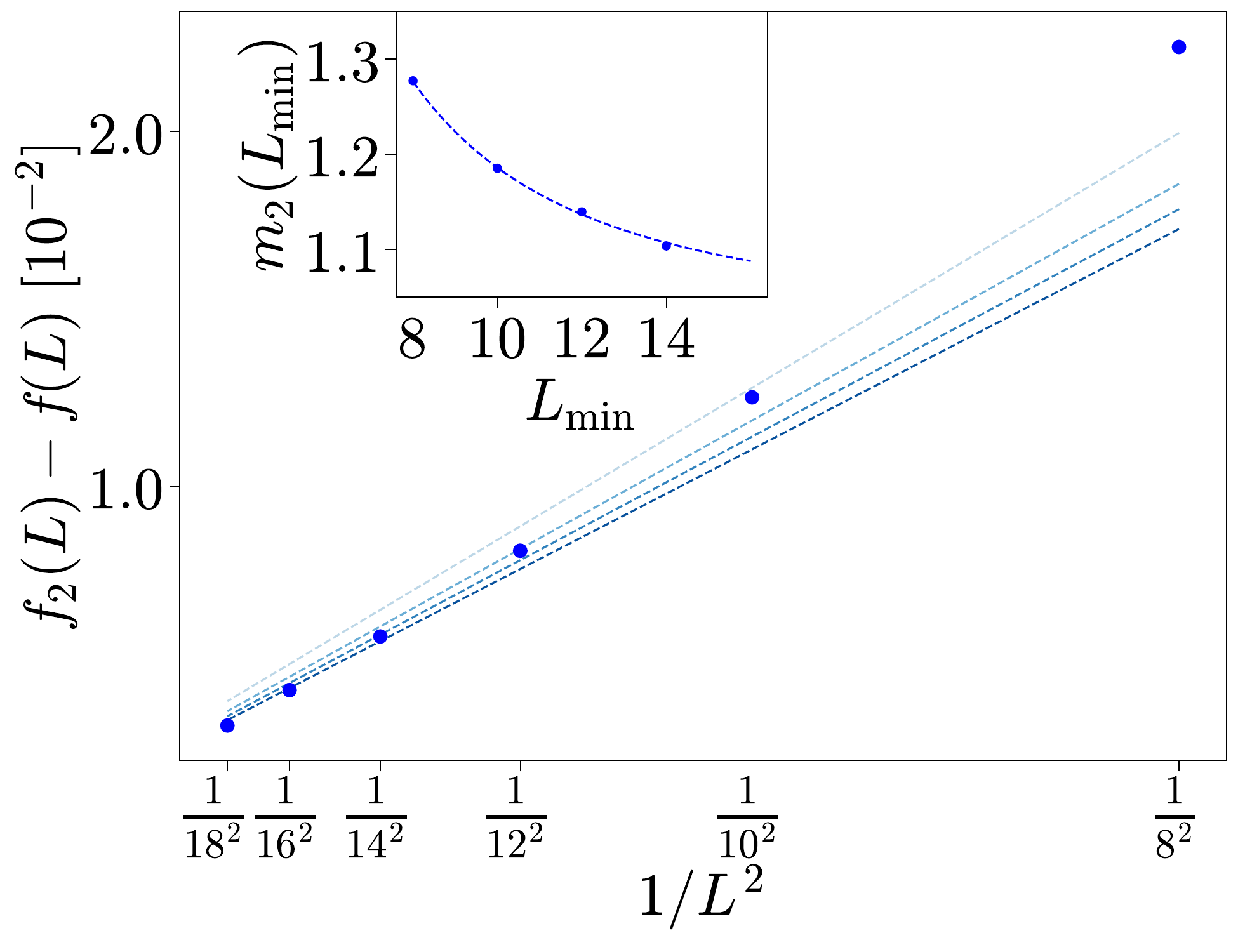}}
    \subfloat[\label{fig:duf3f0}]{\includegraphics[width=0.32\linewidth]{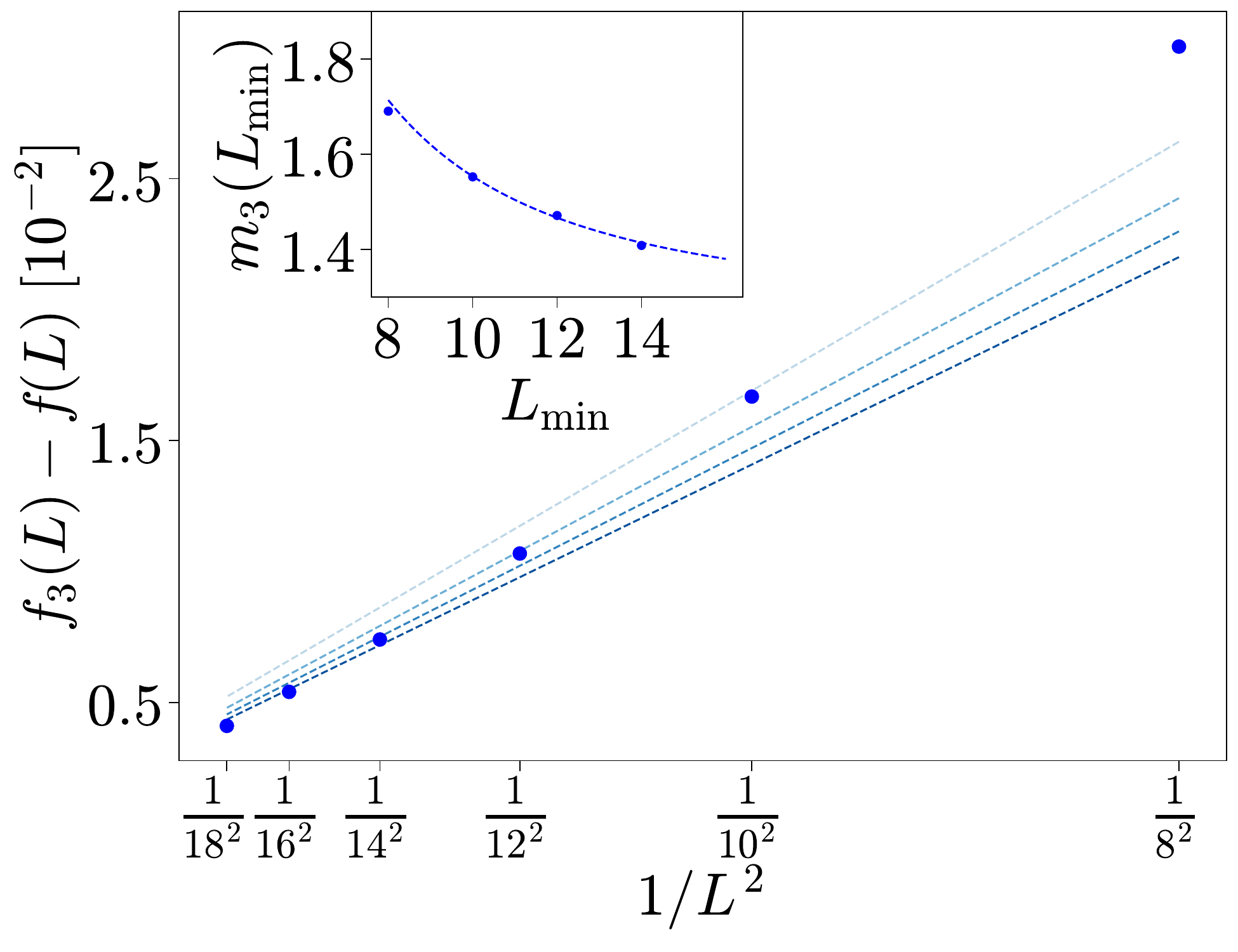}}
    \caption{\label{fig:du_fif0}Scaling of the generalized free energies in the dual unitary model. The differences between the free energy densities
    can be related to the scaling dimension, $x_i^\mathrm{typ}$, of operators in the theory.
    Using a similar double fitting procedure as for the effective central charge, the typical
    values of the first three scaling dimensions are estimated to be
    $x_1^\mathrm{typ} = 0.122(1), x_2^\mathrm{typ} =  0.163(1), x_3^\mathrm{typ} = 0.202(1)$. Darker blue indicates larger
    values of $L_\mathrm{min} = 8\to 14$.}
\end{figure}

The free energy density is shown in the main text where we
extract the effective central charge, $c_\mathrm{eff} = 0.24(2)$. This value
is consistent with the result for the Haar random circuit (see below) but with much smaller
error bars. Additionally, in the main text, we estimated
the scaling dimensions, $x_i^\mathrm{typ}$, of operators in the theory by computing the differences between
the free energy densities. The system size dependence used for the double
fitting procedure is shown in Fig.~\ref{fig:du_fif0} and the equation for each of the fits in the
insets are given by
$m_1(L) = 0.766 + \frac{2.16}{L^2}$,
$m_2(L) = 1.03 + \frac{16.1}{L^2}$,
and $m_3(L) = 1.27 + \frac{28.5}{L^2}$.

\section{Stabilizer circuits}
In this section, we estimate the anisotropy parameter and $c_{\rm eff}$ for the 1+1D random Clifford  model \cite{li2019measurement}.

In the case of a stabilizer circuits, it turns out one can compute the entropy of the measurement record for a fixed circuit without any sampling by simply counting the number of deterministic measurement outcomes $N_{\rm det}$ out of all measurements
\be
F = (N_{\rm meas} - N_{\rm det}) \log 2,
\ee
which follows from the dynamical update rules for stabilizer circuits \cite{Aaronson04}.

\begin{figure}[htbp]
\begin{center}
\includegraphics[width=0.8\textwidth]{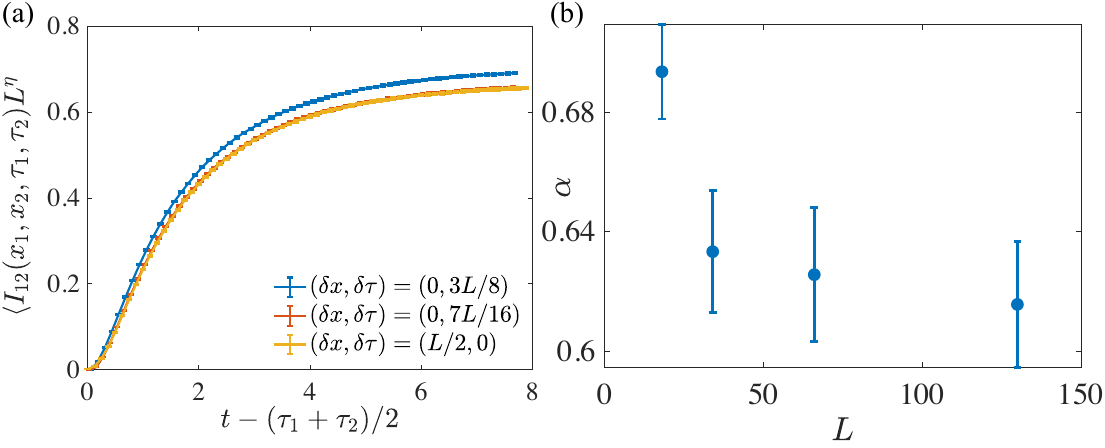}
\caption{(a) Scaled mutual information between two reference qubits for different space and time like separations at $p = 0.1596$ for $L=128$ with the percolation value $\eta = 5/24$.  We averaged over $12 \cdot 10^5$ circuits.   (b) Extracted $\alpha(p,L)$ for $p = 0.1596$ up to $L = 128$.}
\label{fig:anisotropy}
\end{center}
\end{figure}

In Fig.~\ref{fig:anisotropy}, we show numerical data we have used to estimate $\alpha$ up to system sizes $L = 128$.  In Fig.~\ref{fig:anisotropy}a, we show the mutual information between two initially locally entangled reference qubits for space and time-like separations between the reference qubits.

To perform the time-like interpolation we use $\delta_x = 0$ with the separation $\delta \tau = \tau_2 - \tau_1 = 6L/16$ and $7L/16$ that is close to the point where $r = 1$.  As shown in Fig.~\ref{fig:anisotropy}b, at our largest value of $ L = 128$, we find 
\be
\alpha = 0.616 \pm 0.021~ ({\rm stat.}) \pm  0.0037~ ({\rm interp.}) = 0.616 \pm 0.025
\ee
We have estimated the interpolation error arising from a linear interpolation approximation using the formulas
\begin{align}
r(\tau) &\approx 1 - 2 \sqrt{2} \pi \alpha \Delta (\tau-t_*) + \pi^2 \alpha^2 \Delta(1+4 \Delta)(\tau-t_*)^2, \\
\frac{t_*}{L} &= \frac{\tau_1}{L} + \frac{[1-r(\tau_1)] \delta \tau}{[r(\tau_2)-r(\tau_1)] L} + \textrm{Error} \approx 0.4579 \pm 0.0028 ~({\rm interp}),
\end{align}
where we used the estimates $t_* = 0.4579$, $\alpha = 0.616$, and $\Delta = 0.1042$ to approximate the error term arising from the quadratic correction to $r(\tau)$.

The anisotropy parameter as we have defined it will also have corrections due to uncertainty in $p_c$, which leads to the finite size scaling form $\alpha(L/\xi)$, where $\xi \approx X_\pm/|p-p_c|^\nu$.  We previously obtained a quantitative estimate for the prefactor $X_\pm = 0.18/0.12$ above and below the critical point \cite{Gullans20}, which implies that with the currently available precision on $p_c = 0.1593(5)$, the expected correlation length is several hundred to several thousand lattice sites within this uncertainty window.    Numerically, we do not observe any statistically significant dependence of $\alpha(p)$ over this range of $p$.

\begin{figure}[htbp]
\begin{center}
\includegraphics[width=0.8\textwidth]{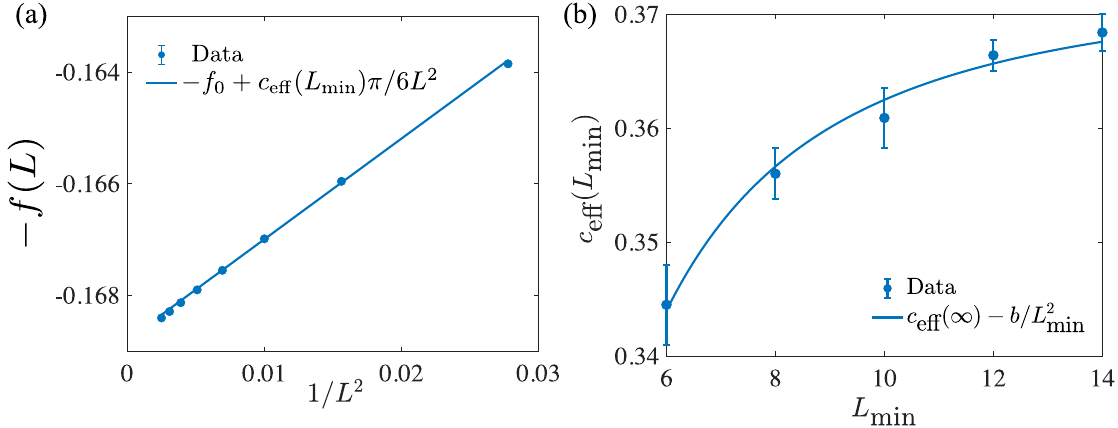}
\caption{(a) Average entropy density of the measurement record vs $1/L^2$ for $L_{\rm min} = 6$.  (b) Dependence of the extracted slope vs $1/L^2$ as a function of the cutoff $L_{\rm min}$.  }
\label{fig:ceff}
\end{center}
\end{figure}

With the anisotropy parameter calibrated, we can now numerically compute the average free energy of the underlying statistical mechanics model.  The numerical results are shown for $p = 0.1596$ in Fig.~\ref{fig:ceff}a, where we see the predicted scaling behavior with $L$. 
By successively removing smaller sizes $L < L_{\rm min}$ from the fit we can obtain a sequence of values $c_{\rm eff}(L_{\rm min})$.  Performing the fit
\be
c_{\rm eff}(L_{\rm min}) = c_{\rm eff}(\infty) - \frac{b}{L_{\rm min}^2},
\ee
allows a reliable method to extract the asymptotic value $c_{\rm eff}(\infty)$ \cite{jacobsen1998critical}. The results of this analysis are shown in Fig.~\ref{fig:ceff}b.  
To determine the variations with $c_{\rm eff}(\infty)$ for different values of $p$ we have scanned several values near the critical point and find the maximum occurs near $ p = 0.1596$, which we use as our estimate of the critical point (see Table~\ref{tab:ceff}).  The variation with $p$ throughout this region is close to the uncertainty in the fits.  

\begin{table}[htp]
\caption{Extracted value of $c_{\rm eff}$ for different values of $p$ near the estimated critical point.}
\begin{center}
\begin{tabular}{c|ccc}
$ p$& 0.1594 & 0.1596 & 0.1598\\
\hline
$c_{\rm eff}(\infty) $ &$ 0.3716 \pm 0.0007$ &$ 0.3729 \pm 0.0016 $&  $0.3705 \pm 0.0009$ \\
\end{tabular}
\end{center}
\label{tab:ceff}
\end{table}%

Overall, we obtain the estimate for $c_{\rm eff}$ including statistical errors and the error in the anisotropy parameter
\be
c_{\rm eff}^s = 0.3729 \pm 0.0016 ~({\rm stat.}) \pm 0.016~({\rm anis.}) = 0.373 \pm 0.018
\ee

\section{Purification exponents in stabilizer circuits and minimal-cut percolation models}

In this section, we compare the order parameter exponent between the minimal-cut/Haar-Hartley percolation universality class and the stabilizer circuit universality class.  We also describe the numerical method we used to more accurately extract the order parameter exponent for the stabilizer circuit models.

\subsection{Purification exponents}

The von Neumann entropy dynamics $S(\rho)$ of a mixed state $\rho = K_{\bm{m}}\rho_0 K_{\bm{m}}^\dag/p_{\bm{m}}$ evolved under a stabilizer circuit has qualitatively the same behavior as the Hartley entropy $S_0(\rho)$ in the Haar random model.  The latter of which has an exact mapping to a percolation problem through the minimal-cut procedure developed in Ref.~\cite{skinner2019measurement}. For this reason, we first benchmark our method on the Haar-Hartley percolation model. 
In both models, the relevant entropy changes in discrete steps of $\log 2$.  At the critical point, we have found that the late time decay rate for the relevant entropy changing from $n \log 2$ to a value $<n \log 2$ saturates to a constant.  This behavior is consistent with a late time exponential decay behavior for the probability of a circuit maintaining entropy $ n \log 2$.  We define the average quantity  
\begin{equation}
    \Delta \lambda_n(t) = \frac{1}{\alpha L} \frac{\textrm{\# Circuits  for which $S(\rho)$ goes from $n \log 2$ to a value $<n \log 2$ at time $t$ }}{\textrm{\# Circuits with $S(\rho)=n \log 2$ at time $t-1$}},
\end{equation}
 Here, $\alpha$ is the anisotropy parameter. In the Haar-Hartley percolation model $\alpha = 1$.  For stabilizer circuits, we focus on the random dual Clifford model where each two-site gate is chosen uniformly randomly from the set of dual-unitary Clifford gates.  This model is expected to have $\alpha = 1$ for each circuit, which we have verified numerically using the method described in the previous section.  This property makes it convenient for numerical analysis similar to the dual-unitary Haar random model. 

To connect this quantity $\Delta \lambda_n(t)$ to more conventional observables at the critical point, we note that, if we start with a mixed state with one bit of entropy, then
\begin{equation}
    \Delta \lambda_1(t) = - \frac{1}{\alpha L \mean{S(\rho)}} \frac{\Delta \mean{S(\rho)}}{\Delta t}, 
\end{equation}
is just the logarithmic time derivative of the entropy of the mixed state.  Within the conformal field theory picture for percolation and the stabilizer circuit models, we have the relation \cite{li2020conformal}
\begin{equation}
    \lim_{t\to \infty} \Delta \lambda_1(t) = \frac{2 \pi}{L^2} x_1,
\end{equation}
where $x_1$ is the order parameter exponent.  Our definition of $\Delta \lambda_n$ allows us to generalize this exponent to an infinite family of ``purification'' exponents.  This spectrum of exponents  serves as a more precise comparison between the stabilzer circuit and  Haar-Hartley percolation universality class.  

Our numerical results for $\lambda_n$ for $n=1 $ and $n=2$  are summarized in Table \ref{tab:xnp}.  For the random Clifford model using these methods,  we find $x_1^p = 0.120(5)$ and $x_2^p=0.240(5)$ with an uncertainty limited mostly by the uncertainty in the anisotropy parameter. In this case, we observe a significant difference from Haar-Hartley percolation values only for $x_2^p$.  On the other hand, for the random dual Clifford model, we observe that it also has a significant difference in the value of $x_1^p$ due to the smaller numerical uncertainties in the estimated value.  This large  relative difference in $x_1^p$ between the two models is a strong indication that they lie in separate universality classes.

\begin{table}[htp]
\caption{The first two purification exponents in the random dual Clifford model and the Haar-Hartley percolation model.  To our knowledge, $x_2^p$ has not been previously studied in percolation.}
\begin{center}
\begin{tabular}{c|c|c|c|c}
&Clifford & Dual Clifford & Haar-Hartley Exact & Haar-Hartley Numerics   \\
\hline
$ x_1^p$& 0.120(5)& 0.111(1)&   $5/48 = 0.1042\ldots$ & 0.104(1)   \\
$x_2^p$ & 0.240(5) &  0.230(1) &   ??? & 0.366(3)  \\
\end{tabular}
\end{center}
\label{tab:xnp}
\end{table}%

\subsection{Numerical method}

\begin{figure}[htbp]
\begin{center}
\includegraphics[width=0.9\textwidth]{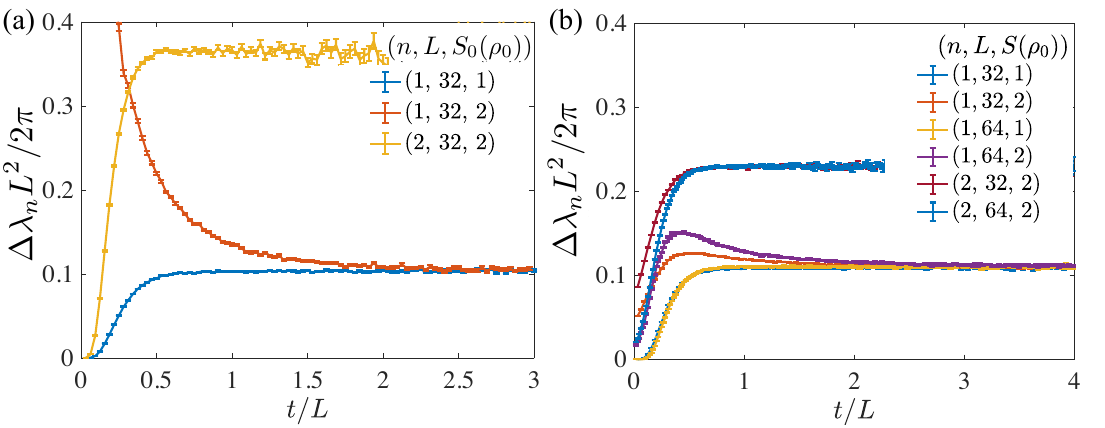}
\caption{(a) Scaled purification rate for the Hartley entropy of the reference system in the Haar random model at the critical point $p= 0.5$.  The decay rate from entropy $n$ to $<n$ allows us to extract the purification exponent $x_n^p$ from the late time plateau.  $x_1^p = x_1^{(1)}$ coincides with the order parameter exponent.  (b) Scaled purification rates for the entropy of the reference system in the random dual  Clifford model at $p = 0.205 \approx p_c$.  }
\label{fig:xnp}
\end{center}
\end{figure}

Our numerical method used for extracting the purification exponents is illustrated in Fig.~\ref{fig:xnp} for the Haar-Hartley percolation model and the random dual Clifford model.
To improve the numerical precision for $x_1^p$, we choose different initial conditions whereby the decay rate approaches its late time plateau from either above or below the plateau value.  By averaging these two results, we can reduce systematic errors in our numerical estimate of the plateau value.  

For the Haar-Hartley percolation model shown in Fig.~\ref{fig:xnp}a, we took an initial state with Hartley entropy $S_0(\rho) = 2$ or 1, fully scrambled the system with a Haar random circuit, and then turned on the measurements at the critical rate $p=0.5$.  In the percolation mapping, the scrambling layer corresponds to taking a fully connected bottom boundary. To compute $S_0(\rho)$, we used the max-flow/min-cut algorithm applied to a percolating network.  With this method, we were able to extract a value of $x_1^p$ that is with $1 \% $ of the known percolation value of $5/48$.  To our knowledge, the exact values of $x_n^p$ for $n>1$ are not known within the minimal cut picture for the Haar-Hartley entropy.  We provide the first  numerical estimate of $x_2^p$ here.

For the random dual Clifford model shown in Fig. \ref{fig:xnp}b,   the boundary conditions were chosen in a similar manner to the Haar-Hartley model; however, to improve the rate at which the $S(\rho_0) = 2$ initial condition approaches the plateau, we scrambled the initial condition with a depth $L$ random circuit that also includes  measurements at rate $p = p_c/1.25$.  As a result, the quench to the critical point is less dramatic compared to a fully unitary scrambling circuit.  For the initial condition $S(\rho_0) = 1$, the scrambling layer was taken to be a depth $2L$ random  Clifford circuit in 1D with no measurements.  The critical point $p_c = 0.205(1)$ for the random dual Clifford model was obtained using the order parameter crossing method described in our previous work \cite{gullans2020scalable}. The extracted value of $p_c$ strongly violates the Hashing bound for a depolarizing channel  that was conjectured to be a relevant bound on the critical measurement rate $p_c \le 0.1893$ for unitary-projective circuits in one dimension \cite{Fan20}.

\section{Effective central charge in the large onsite Hilbert space dimension limit}

In this section, we derive exact expressions for the effective central charge $c_{\rm eff}$ of the MIPT of monitored qudit circuits for both Haar and Clifford random gates, in the limit $d \to \infty$ where $d$ is the dimension of the onsite Hilbert space. Note 
that, as already recalled in a footnote in the introductory part of the main text, $c_{\rm eff}$ is {\em not} related to the prefactor of the logarithmic scaling 
with subsystem size of the entanglement entropy at criticality, which is instead related to the scaling dimension of boundary condition changing operators~\cite{RTN2,jian2020measurement}. 

\subsection{Haar case}

In the case of Haar gates drawn from the unitary group $U(D=d^2)$,  we follow Refs.~\cite{bao2020theory,jian2020measurement} (see also~\cite{RTN1, RTN2, PhysRevB.99.174205}) to map the anneal averaged replicated partition functions $\bar{Z}_r = \sum_{\mathbf{m}} p_{\mathbf{m}}Z_{\mathbf{m}}^r$ onto an effective statistical model (recall, $Z_{\mathbf{m}} = p_{\mathbf{m}}$ in our formulation), whose degrees of freedom are permutations $g_i \in S_{1+r}$. Formally, this follows from the so-called Schur-Weyl duality, which states that the permutation group $S_{1+r}$ and the unitary group $U(D)$ act on $({\mathbb C}^D)^{\otimes (1+r)}$ as a commuting pair. In the limit $d \to \infty$, the statistical mechanics model simplifies dramatically, and reduces to a Potts model with $\left| S_{r+1} \right| = (r+1)!$ states. In the replica limit $r \to 0$, this gives a MIPT in the percolation universality class~\cite{bao2020theory,jian2020measurement}. 

For a finite number of replicas $r$, this Potts model has a phase transition described by a CFT with central charge 
\begin{equation}
c(r) = 1 - \frac{6}{x(x+1)} \ {\rm with} \ x+1 = \frac{\pi}{\arccos \frac{\sqrt{(r+1)!}}{2}}.
\end{equation}
In the replica limit, we have $c(r \to 0) = 0$, and we can use this expression to evaluate the effective central charge
\begin{equation}
c^{H,d \to \infty}_{\rm eff} = \lim_{r \to 0} \frac{d c}{dr}  = \frac{5 \sqrt{3} (1 - \gamma)}{4 \pi} \simeq 0.291367\dots
\end{equation}
with $\gamma \simeq 0.577216\dots$ Euler's constant. 

\subsection{Clifford case}

We now turn to a similar calculation in the case of Clifford gates. The full derivation of the corresponding statistical mechanics model (for the Clifford measurement-induced phase transition and random tensor networks~\cite{RTN2} with Clifford tensors) with on-site Hilbert space dimension $d=p^n$ and $p$ prime will be reported elsewhere~\cite{UnpublishedClifford}, where it will also be shown that its symmetry depends explicitly on $p$, implying universality of transitions depending on $p$. Here we simply emphasize the key ingredients to compute $c_{\rm eff}$ in the limit of large onsite Hilbert space.  In order to average over Clifford gates to derive a statistical model, we will need a generalization of the Schur-Weyl duality. Let $D=d^2$ with $d=p^n$ and $p$ prime. We are interested in the Clifford group ${\cal C}$, which is a finite subgroup of the unitary group $U(D)$ acting on $r+1$ replicas.  In general, the ``commutant'' ${\cal F}_{r}$ of  ${\cal C}$ acting on this space will be larger than the symmetric group $S_{r+1}$, and was recently analyzed in  Ref.~\cite{2017arXiv171208628G}. 
In order to analyze the structure of this algebraic object, note that the tensor space $V=({\mathbb C}^D)^{\otimes (r+1)}$ can be decomposed onto the irreps $V_\lambda$ of ${\cal C}$ as $V= \bigoplus_\lambda d_\lambda V_\lambda$.
The dimension of the commutant ${\cal F}_{r+1}$  of the Clifford group ${\cal C}$ acting on this replicated space is $|{\cal F}_{r+1}| = \sum_\lambda d_\lambda^2$, and can be computed as follows. Let $\chi_{V}(U) = {\rm tr} \ U^{\otimes (1+r)} $ be the character of the representation $({\mathbb C}^D)^{\otimes (1+r)}$ of  the Clifford group ${\cal C}$, where $U \in {\cal C}$ is a Clifford gate acting  on ${\mathbb C}^D$. Introducing the inner product between characters $\langle \chi_1 , \chi_2 \rangle = \frac{1}{|{\cal C}|} \sum_{U \in {\cal C}} \overline{\chi_1(g)} \chi_2(g)$, we have $\langle \chi_{V} , \chi_{V} \rangle = \sum_\lambda d_\lambda^2$. The dimension of the commutant ${\cal F}_{r+1}$ of the Clifford group -- which replaces the symmetric group $S_{r+1}$ in the statistical mechanics model -- is thus given by
\begin{equation}
| {\cal F}_{r+1} |= \frac{1}{|{\cal C}|} \sum_{U \in {\cal C}} \left| {\rm tr} \ U \right|^{2(r+1)}.
\end{equation}
This quantity is known as a ``frame potential'' in the quantum information literature. In general, the structure of $ {\cal F}_Q$ will depend on $d=p^n$. If we focus on $d=2^n$ with large $n$ (we will report on the other cases elsewhere~\cite{UnpublishedClifford}), the dimension of the commutant saturates with $n$ to a quantity strictly larger than $(r+1)!$~\cite{2017arXiv171208628G}
\begin{equation}
|{\cal F}_{r+1}| = \prod_{k=0}^{r-1} \left(2^k+1 \right)= 2^{r(r-1)/2} \frac{ \displaystyle\prod_{k=0}^{\infty} \left(1 + \frac{1}{2^k} \right)}{ \displaystyle\prod_{k=0}^{\infty} \left(1 + \frac{1}{2^{k+r}} \right)}, 
\end{equation}
where $r$ can be analytically continued to be a real number in the right-hand side. 
The statistical mechanics model of monitored Clifford circuits will involve degrees of freedom living in $  {\cal F}_{r+1}$, which in general has a complicated algebraic structure~\cite{2017arXiv171208628G}, not relevant to us here. In the limit $n \to \infty$, we expect that the statistical mechanics model reduces once again to a Potts model with $|{\cal F}_{r+1}| $ states: this is because any generalization of the Weingarten functions of Haar calculus will become proportional to delta functions in that limit. This is a large $D$ limit, as in the Haar case (except there are different ways to approach this limit in the Clifford case, here we set $D=p^{2n}$ and took $n \to \infty$ with $p=2$). The central charge as a function of the number of replicas $r$ is now given by $c(r) = 1 - \frac{6}{x(x+1)}$ with $ x+1 =\frac{\pi}{ \arccos \frac{\sqrt{|{\cal F}_{r+1}|}}{2}}$.
This leads to a closed form expression for the effective central charge 
\begin{equation}
c^{C, d=2^n \to \infty}_{\rm eff} = \frac{5 \sqrt{3}}{8 \pi} \left( 2 \psi_{\frac{1}{2}}\left( \frac{- i \pi}{\log 2}\right) - \log 8 \right) \simeq 0.365194\dots
\end{equation}
where $\psi_q(z)$ is the $q$-digamma function, which is defined as the derivative of $\log \Gamma_q(z)$ with respect to $z$, where  $\Gamma_q(z)$ is the $q$-deformed Gamma function. 

\end{document}